\documentclass[apj,numberedappendix]{emulateapj}
\usepackage[T1]{fontenc}
\usepackage[latin9]{inputenc}
\usepackage{amssymb}
\usepackage{graphicx}
\usepackage[unicode=true,pdfusetitle,
 bookmarks=true,bookmarksnumbered=false,bookmarksopen=false,
 breaklinks=false,pdfborder={0 0 1},colorlinks=false]
 {hyperref}

\makeatletter
\providecommand{\tabularnewline}{\\}

\usepackage{amsmath}

\makeatother

\begin{document}

\global\long\def\De{\Delta\epsilon}
\global\long\def\Demax{\Delta\epsilon_{\max}}
\global\long\def\Demin{\Delta\epsilon_{\min}}

\global\long\def\DE{\Delta E}
\global\long\def\DEmax{\Delta E_{\max}}
\global\long\def\DEmin{\Delta E_{\min}}

\global\long\def\DtE{\Delta_{t}E}

\global\long\def\Dv{\Delta v}

\global\long\def\Dvvec{\Delta\mathbf{v}}

\global\long\def\vvec{\mathbf{v}}

\global\long\def\xvec{\mathbf{x}}

\global\long\def\Dt{\Delta t}

\title{Stellar Energy Relaxation around A Massive Black Hole}

\author{Ben Bar-Or, G\'{a}bor Kupi and Tal Alexander}

\affil{Department of Particle Physics \& Astrophysics, Faculty of Physics,
Weizmann Institute of Science, POB 26, Rehovot, Israel}
\begin{abstract}
Orbital energy relaxation around a massive black hole (MBH) plays
a key role in establishing the stellar dynamical state of galactic
nuclei, and the nature of close stellar interactions with the MBH.
The standard description of this process as diffusion in phase space
provides a perturbative 2nd-order solution in the weak two-body interaction
limit. We carry out a suite of $N$-body simulations, and find that
this solution fails to describe the non-Gaussian short timescale evolution
of the energy, which is strongly influenced by extreme events (a ``heavy-tailed''
distribution with diverging moments) even in the weak limit, and is
thus difficult to characterize and measure reliably. We address this
problem by deriving a non-perturbative solution for energy relaxation
as an anomalous diffusion process driven by two-body interactions,
and by developing a robust estimation technique to measure it in $N$-body
simulations. These make it possible to analyze and fully model our
numerical results, and thus empirically validate in detail, for the
first time, this theoretical framework for describing energy relaxation
around an MBH on all timescales. We derive the relation between the
energy diffusion time, $t_{E}$, and the time for a small density
perturbation to return to steady state, $t_{r}$, in a relaxed, single
mass $n(r)\propto r^{-7/4}$ cusp around a MBH. We constrain the modest
contribution from strong stellar encounters, and measure with high
precision that of the weakest encounters, thereby determining the
value of the Coulomb logarithm, and providing a robust analytical
estimate for $t_{E}$ in a finite nuclear stellar cusp\@. We find
that $t_{r}\simeq10t_{E}\simeq(5/32)Q^{2}P_{h}/N_{h}\log Q$, where
$Q=M_{\bullet}/M_{\star}$ is the MBH to star mass ratio, the orbital
period $P_{h}$ and number of stars $N_{h}$ are evaluated at the
energy scale corresponding to the MBH's sphere of influence, $E_{h}=\sigma_{\infty}^{2}$,
where $\sigma_{\infty}$ is the velocity dispersion far from the MBH\@.
We conclude, scaling $\sigma_{\infty}$ by the observed cosmic $M_{\bullet}/\sigma_{\infty}$
correlation, that stellar cusps around lower-mass MBHs ($M_{\bullet}\lesssim10^{7}M_{\odot}$),
which have evolved passively over a Hubble time, should be dynamically
relaxed. We briefly consider the implications of anomalous energy
diffusion for orbital perturbations of stars observed very near the
Galactic MBH. 
\end{abstract}


\section{Introduction}

\subsection{Background}

Observations of our own Galactic Center (GC) reveal that stars there
move on Keplerian orbits in the potential of a central compact object,
which is consistent with a massive black hole (MBH) of mass $M_{\bullet}\sim4\times10^{6}M_{\odot}$
\citep{Eisenhauer2005,Ghez2005,Ghez2008,Gillessen2009a,Gillessen2009}.
It is widely assumed that the Milky Way is not unique in this sense,
and that there is an MBH in the center of most, if not all, galaxies.

The GC, with its relatively low-mass MBH, represents a sub-class of
galactic nuclei where the dynamical relaxation time is expected to
be shorter than the Hubble time \citep{Alexander2007a}, so that the
stellar distribution near the MBH could have had time to reach a relaxed
state that is independent of initial conditions. Moreover, because
of its proximity, it is the only system to date where the dynamical
state can be directly probed by observations. However, recent star
counts in the GC \citep{Buchholz2009,Do2009,Bartko2010} indicate
that the radial density profile of the spectroscopically identified
low-mass red giants on the $\sim0.5$ pc scale, thought to trace the
long-lived population there, rises inward much less steeply (or even
decreases) than is expected for a dynamically relaxed stellar population
\citep{Bahcall1976,Bahcall1977,Alexander2009,
Preto2010}. The simplest interpretation is that, contrary to expectations, the
GC is not dynamically relaxed by two-body interactions. This could
be due, for example, to a cosmologically recent major merger event
with another MBH, which carved out a central cavity in the stellar
distribution \citep{Merritt2010b}, or due to the presence of faster
competing dynamical processes that drain stars into the MBH \citep{Madigan2011}.

This interpretation is however further complicated by a recent study
of the diffuse light density distribution in the GC \citep{Yusef-Zadeh2012},
which is thought to track the low-mass, long-lived main-sequence stars
that are too faint and numerous to be individually resolved, and consist
of the bulk of the stellar mass. In contrast with the stellar number
counts, the diffuse light distribution appears to rise inward rather
steeply down to the $\sim O(0.01)$ pc scale, where destructive stellar
collisions (not accounted for in the purely gravitational treatment
of two-body relaxation), can plausibly suppress the stellar density
profile \citep{Alexander1999a}. 

These uncertainties about the dynamical state of the GC, and by implication,
the state of galactic nuclei in general, have far-reaching ramifications.
The possibility that the GC is unrelaxed severely limits the validity
of extrapolating results derived from the unique observations of the
GC to other galactic nuclei. While a relaxed system can be understood
and modeled from first principles, independently of initial conditions,
an unrelaxed one reflects its particular formation history, which
is expected to vary substantially from galaxy to galaxy. This, and
the possibility that the GC lacks the predicted high-density central
stellar cusp, have in particular crucial implications for attempts
to understand the dynamics of extra-galactic gravitational wave (GW)
sources and to predict their rates, since the low-mass MBH in the
GC has long been considered the archetype of low-frequency extra-galactic
targets for planned space-borne GW observatories \citep{Amaro-Seoane2007}.

The study of dynamical relaxation around a MBH has a long history,
beginning with the seminal studies of \citet{Peebles1972,Bahcall1976,Bahcall1977,Cohn1978,Shapiro1978},
who used various simplifying approximations \citep{Nelson1999} to
enable analytical and numerical approaches, and continuing more recently
with studies of the time to approach steady state near a MBH in large
$N$-body simulations \citep[e.g.][]{PretoMiguel2004,Preto2010}.
A complementary approach of extracting the diffusion coefficients
(DCs) from stellar energy evolution in $N$-body simulations was carried
out in several studies \citep{Lin1980,Rauch1996,Eilon2009,Madigan2011},
leading to widely differing predictions about the dynamical state
of galactic nuclei, especially very close to the MBH \citep{Madigan2011}.
These open questions, and the fundamental importance of the process
of orbital energy relaxation, motivate a re-examination of the analytic
description of energy relaxation, coupled to detailed comparisons
with carefully controlled simulations.

\subsection{Objectives and overview of this study}

In this study we use analytical considerations and high-accuracy $N$-body
simulations to address the following open questions: What is the relation
between local energy relaxation and the approach to a global steady
state? What are the properties of evolution due to energy relaxation
on the short and intermediate timescales? How rare are extreme energy
exchange events, and what is their contribution to relaxation? What
is the physical meaning, and the correct value, of the Coulomb cutoffs
introduced to control the formal divergences in the relaxation rates?
In practical terms, these questions all reduce to the problem of relating
the analytical / statistical description of local two-body interactions,
to the global measurable properties of a stellar system. These issues
must be addressed in order to better understand dynamical processes
in galactic nuclei, improve their modeling, and reliably apply results
from $N$-body simulations to real physical systems.

Our approach to these problems is complementary to that of \citet{PretoMiguel2004}.
\citet{PretoMiguel2004} simulated a very large system with a non-Keplerian,
and non-self-similar initial DF and potential and a relatively low
mass ratio $Q=M_{\bullet}/M_{\star}$ over a relaxation timescale.
\citet{PretoMiguel2004} show that there is an agreement between the
$N$-body and Fokker-Planck (FP) simulations by comparing the evolution
of a small inner sub-system where the potential is dominated by that
of the MBH and where the analytical/ numerical solution of \citet{Bahcall1976}
(henceforth the BW cusp) is relevant. 

Here, we carry out a suite of small $N$-body simulations of a system
with realistically high values of $Q$ and a potential that is dominated
by the MBH, designed specifically to simplify the analytic treatment
and to allow complete modeling of the time-dependent evolution on
short timescales. This allows us to rigorously test and verify the
theoretical predictions, and in particular the scaling of the relaxation
with energy and with the slope of the cusp.

The choice of focusing on short timescales is to a large extent forced
by the high computational cost of the simulations, since we model
a high mass ratio system ($Q=10^{6}$, comparable to the mass ratio
in the GC) with $N\ge10^{4}$ particles which are strongly bound to
the MBH. The short timescale behavior is however also of interest
in itself, since dynamical processes near a MBH can be limited by
the lifetime of the stars (for example, the massive stars in the GC),
or by fast destructive processes (for example, collisional destruction
or tidal disruption).

The elementary description of energy relaxation is by the energy \emph{transition
probability}, $K_{E,J}\left(\DE\right)$ (Eq.~\ref{eq:K-JE}), which
is the orbit-averaged probability per unit time, per unit energy,
for the energy of a star to change from $E$ to $E+\Delta E$ in a
single scattering event. The resulting distribution of the accumulated
change in energy over a finite time $t$, $\Delta_{t}E$, is represented
by the \emph{propagator}, $W_{E,J}(\Delta_{t}E,t)$, which describes
the energy distribution of an initially mono-energetic system of test
stars, after a time $t$.

Past standard treatments of energy relaxation in stellar systems \citep[e.g.][]{Bahcall1976}
with the FP formalism implicitly assumed that only the two lowest
moments of the energy transition probability (i.e. the 1st and 2nd
DCs) play part in the evolution of the system. This is equivalent
to assuming that the free propagator is a Gaussian with width proportional
to $\sqrt{t}$, that is, energy evolves by normal diffusion. However,
as we show, the $\sim1/|\DE|^{3}$ divergence of the transition probability
\citep{Goodman1983} implies that higher order DCs do play a role,
which is dominant on short timescales, and diminishes with time, but
still remains non-negligible even when the system relaxes to steady
state. The implications are that energy relaxation progresses somewhat
faster than $\sqrt{t}$, as an anomalous diffusion process.

Our approach of measuring energy relaxation in $N$-body simulations
on short timescales forces us to develop a robust statistical model
for describing these timescales. By treating the problem in Fourier
space, we are able to derive the full propagator explicitly. It can
then be evaluated either numerically in non-perturbative form, or
analytically in a perturbative manner. This enables us to show that
the propagator is a heavy-tailed DF that evolves as $\sim\sqrt{t\log(t)}$
on timescales much shorter than the FP energy diffusion timescale.
This short timescale behavior is an example of an anomalous diffusion
process \citep{Metzler2000}. One consequence is the much higher probability
of events that would be expected to be extremely rare for a Gaussian
probability distribution. Another implication of the very slow convergence
of this propagator to the central limit is that attempts to measure
relaxation in relatively short $N$-body experiments can yield very
misleading results, which reflect more the method used, than the quantity
probed. By analyzing the scaling properties of the energy propagator
we suggest a robust statistical estimator, using fractional moments,
which is independent of the high energy cutoff. With it we measure
the rate of energy relaxation on short timescales and confirm our
analytical treatment, and in particular, the validity of the propagator
and the DCs used in the standard FP approximation. Finally, we use
the propagator to evolve a system by MC simulations up to steady state,
and thereby confirm the approximate validity of the FP results on
long timescales.

The paper is organized as follows: In Section~\ref{s:E-relax} we
summarize the process of energy relaxation around a massive object
in terms of a normal diffusion process described by the FP equation,
and derive an analytical solution for the evolution of a stellar cusp
to steady state. In Section~\ref{s:beyondFP} we present an elementary
description of energy relaxation in terms of the transition probability
and its integrated propagator, and describe the emergence of anomalous
diffusion that ultimately approaches normal diffusion on long timescales.
In section~\ref{s:measurements} we formulate a method to analyze
our $N$-body simulations and use it to confirm and calibrate our
analytical description of relaxation. In section~\ref{s:GCorbit}
we briefly discuss the observable effects of anomalous diffusion on
short period stars orbiting the Galactic MBH, perturbed by a dark
cusp of stellar remnants. We summarize and discuss our results in
section~\ref{s:summary}.

\section{Energy relaxation as a diffusion process}

\label{s:E-relax}

The standard approach \citep{Chandrasekhar1943} of treating relaxation
in a self-gravitating system involves several simplifying assumptions
\citep[e.g.][]{Nelson1999}. The 2-body perturbations experienced
by a test star are assumed to be independent of each other, that is,
non-coherent. The encounter is assumed to be between two unbound stars,
approaching each other from infinity. The perturbations are assumed
to be local in two respects. (1) The exchange of energy and angular
momentum occurs over a very short time at the point of closest approach.
(2) The local conditions (stellar density and velocity) at the position
of the test star, at the time of interaction, apply to all perturbers
irrespective of their impact parameter, which is equivalent to assuming
an infinite homogeneous system around the point of interaction. These
assumptions cannot be satisfied for arbitrarily large impact parameters,
and therefore require the introduction of a large-distance cutoff,
which enters in the Coulomb factor $\Lambda$. Note however that it
is far from obvious that these assumptions hold near a MBH. The potential
there is dominated by the central mass, and on intermediate scales
(close to the MBH, but far enough to ignore relativistic effects)
is approximately Keplerian, $\phi=GM_{\bullet}/r$; the stellar density
decreases steeply with radius; and there is strong energy-dependent
stratification of dynamical timescales. We discuss here the implications
of these factors on energy relaxation near a MBH. Finally, it is commonly
assumed that this Markovian process is also a diffusive processes,
and can therefore described by the FP equation.

\subsection{The FP description of relaxation around an MBH}

Determining the stellar distribution in the vicinity of an MBH is
a classical problem in stellar dynamics, first discussed in the context
of an MBH embedded in a globular cluster \citep{Peebles1972,Bahcall1976}.
The general time evolution of the phase-space distribution function
(DF) of test particles with mass $m_{T}$, $f_{T}(\vvec,\xvec,t)$,
due to two-body interaction with field stars in a local velocity-mass
distribution $f_{F}(\vvec,m_{F})$, is given by the master equation
\citep{GoodmanJ.1985} 
\begin{eqnarray}
\frac{D}{Dt}f_{T}(\vvec,\xvec,t) & = & \int d^{3}\vvec^{\prime}f_{T}(\vvec^{\prime},\xvec,t)K(\vvec^{\prime},\vvec)\nonumber \\
 &  & -f_{T}(\vvec,\xvec,t)\int d^{3}\vvec^{\prime}K(\vvec,\vvec^{\prime})\,,\label{eq:Velocity-Master-eq}
\end{eqnarray}
where $D/Dt\equiv(\partial/\partial t)+\vvec\cdot(\partial/\partial\xvec)-(\partial\phi/\partial\xvec)\cdot(\partial/\partial\vvec)$
and $K(\vvec,\vvec^{\prime})$ is the transition probability per unit
time per unit velocity volume that a test star changed its velocity
from $\vvec$ to $\vvec^{\prime}=\vvec+\Dvvec$ as a result of a single
encounter with a field star 
\begin{eqnarray}
K(\vvec,\vvec^{\prime}) & = & \frac{4G^{2}}{|\Dv|^{5}}\int dm_{F}m_{F}^{2}\int_{\mathbf{w}\cdot\Dvvec}d^{2}\mathbf{w}\nonumber \\
 &  & f_{F}\left(\mathbf{w}+\vvec+\Dvvec\frac{m_{T}+m_{F}}{2m_{F}},m_{F}\right)\,.\label{eq:Kvv}
\end{eqnarray}

This description of stellar dynamics as a relaxation process assumes
that changes in energy and angular momentum result from uncorrelated
velocity deflections, which are local in time and space \citep[e.g.][]{Nelson1999,Freitag2008}.
Dynamical mechanisms that involve long timescale correlations, e.g.
resonant relaxation \citep{Rauch1996}, are not taken into account
here.

Assume that $f_{T}$ and $f_{F}$ are functions only of orbital energy
$E$ and time $t$; the potential is dominated by a massive central
object $\phi=GM_{\bullet}/r$; and the field stars and the test stars
have the same mass, $M_{\star}$. The master equation \eqref{eq:Velocity-Master-eq}
is then reduced to an equation of energy only (here and below the
energy and potential of bound stars is defined as positive),
\begin{eqnarray}
\frac{\partial N(E,t)}{\partial t} & = & \int d\DE\{N(E-\DE)K_{E-\DE}(\DE)\nonumber \\
 &  & -N(E)K_{E}(\DE)\}\,,\label{eq:M-equ-E}
\end{eqnarray}
where $K_{E}\left(\De\right)$ is the orbit- and eccentricity-averaged
energy transition probability, and $N(E)$ is the stellar number density
in energy space

\begin{equation}
N(E,t)=4\pi^{2}J_{c}^{2}(E)P(E)f(E,t)\,.\label{eq:N(E)}
\end{equation}

By assuming weak encounters, $\DE\ll E$, and expanding $N\left(E-\DE\right)$
and $K_{E-\DE}\left(\DE\right)$ around $E-\DE=E$, the master equation
can be expressed as a Kramers-Moyal expansion \citep[e.g.][]{Weinberg2001}
\begin{equation}
\frac{\partial N(E,t)}{\partial t}=\sum\frac{1}{n!}\left(-\frac{\partial}{\partial E}\right)^{n}N(E)\Gamma_{n}^{\Lambda}\left(E\right)\,,
\end{equation}
where the DCs are the moments of the transition probability $K_{E}\left(\DE\right)$,
\begin{equation}
\Gamma_{n}^{\Lambda}\left(E\right)=\int_{\DEmin}^{\DEmax}\DE^{n}K_{E}\left(\DE\right)d\DE\,,
\end{equation}
and where $\Lambda=\DEmax/\DEmin$. For $n=1,2$, $\Gamma_{n}^{\Lambda}$
is proportional to the Coulomb logarithm $\log\Lambda$ (Eqs. \ref{eq:Gamma1-cusp-inf},
\ref{eq:Gamma2-cusp-inf}) ; for higher moments the dependence is
much weaker, $\propto(1-\Lambda^{-\left\lfloor (n-1)/2\right\rfloor })$
(Eq. \ref{eq:DCsH}), and can usually be ignored.

By truncating the sum after the first two terms, the Kramers-Moyal
expansion is reduced to the FP equation \citep[e.g.][]{Rosenbluth1957}
\begin{eqnarray}
\frac{dN(E,t)}{dt} & = & \frac{1}{2}\frac{\partial^{2}}{\partial E^{2}}\{N(E,t)\Gamma_{2}^{\Lambda}(E)\}\nonumber \\
 &  & -\frac{\partial}{\partial E}\{N(E,t)\Gamma_{1}^{\Lambda}(E)\}=-\frac{\partial F(E)}{\partial E}\,,\label{eq:FP-E}
\end{eqnarray}
where $F(E)$ is the stellar flux.

The FP approximation is valid only if the higher order terms can be
neglected. This is usually justified by arguing that the $n>2$ terms
of the expansion are of the order of the $n>2$ DCs themselves, which
are smaller than the first two DCs by a factor of $1/\log\Lambda$
\citep[e.g.][]{BinneyJames2008, Henon1960}. However, as we show in
Section~\ref{sub:st-propagation}, such a comparison is meaningful
only when $\partial/\partial E\sim1/\Delta E\sim1/E$ (Eq. \ref{eq:DCsH}),
i.e. the change in energy has accumulated to an order unity change.
This is a valid assumption only over long timescales; on short timescales
the system cannot be described by the first two DCs only.

It is further commonly assumed that stars that are unbound to the
MBH ($E<0$) are drawn from a Maxwell-Boltzmann distribution, 
\begin{equation}
f(E)=\frac{n_{0}}{\left(2\pi\sigma_{0}^{2}\right)^{3/2}}e^{E/\sigma_{0}^{2}},\quad E<0\,,\label{eq:boundary-out}
\end{equation}
where $n_{0}$ and $\sigma_{0}$ are the stellar number density and
velocity dispersion of this simplified model of the host galaxy. The
inner energy boundary is determined by the maximal orbital energy
$E_{d}$ where stars can survive against any of the various disruption
mechanisms that operate in the extreme environment near a MBH (e.g.
tidal disruption, tidal heating, orbital decay by gravitational wave
emission, stellar collisions), 
\begin{equation}
f(E)=0,\quad E<E_{d}\,.\label{eq:boundary-in}
\end{equation}

The steady state solution ($\partial N/\partial t=0$) of the FP equation
can be simply obtained if a cusp solution $f(E)=CE^{p}$ is assumed
\citep[e.g.][]{Peebles1972}. Although this solution does not satisfy
the boundary conditions (Eqs.~\ref{eq:boundary-out}, \ref{eq:boundary-in}),
it is in fact a reasonable approximation of the full solution far
from the boundaries at $0\ll E\ll E_{d}$, \citep[e.g.][]{Bahcall1976}.
For a full analytic solution see \citet{Keshet2009b}.

Approximating the DF by a pure power-law, $-1<p<1/2$, the DCs are
\begin{equation}
\Gamma_{1}^{\Lambda}(E)=-\frac{4(1-4p)\log\Lambda}{1+p}\frac{N(<E)}{Q^{2}P(E)}E\propto E^{p+1}\label{eq:Gamma1-cusp-inf}
\end{equation}
and 
\begin{equation}
\Gamma_{2}^{\Lambda}(E)=\frac{16\left(3-2p\right)\log\Lambda}{(1+p)(1-2p)}\frac{N(<E)}{Q^{2}P(E)}E^{2}\propto E^{p+2}\,.\label{eq:Gamma2-cusp-inf}
\end{equation}
The steady state solution is then (Eq.~\ref{eq:FP-E}) 
\begin{equation}
(4p-1)(4p-3)\frac{2}{1-2p}=0\,,\label{eq:SS-solution}
\end{equation}
whose solutions are either $p=3/4$ \citep{Peebles1972}, which is
unphysical since $\Gamma_{2}^{\Lambda}(E)$ diverges with $E_{d}\to\infty$
leading to huge stellar flux away from the MBH \citep{Bahcall1976},
or $p=1/4$, the so-called BW cusp solution \citep{Bahcall1976}.
The BW cusp solution has zero net flow ($F=0$) since the term in
the equation that contains the second DC is independent of energy,
and since there is no drift \citep{Lightman1977}. This solution
was confirmed by integrating the FP equation, both numerically \citep[e.g.][]{Bahcall1976, Cohn1978, PretoMiguel2004}
and statistically \citep[e.g.][]{Shapiro1978}, by Monte-Carlo (MC)
simulations.

\subsection{Relaxation to steady state}

\label{s:relaxation-times}

Stellar systems out of equilibrium evolve by redistributing their
energy and angular momentum. When this proceeds by diffusion, the
timescales associated with the various DCs generally reflect the dynamical
evolution timescale. One such timescale is the energy diffusion time
$t_{E}$, 
\begin{equation}
t_{E}(E)\equiv E^{2}/\Gamma_{2}^{\Lambda}(E)\label{eq:tE-E}
\end{equation}
which expresses the timescale for a test star to change its orbital
energy by order unity. Another commonly used quantity is the Chandrasekhar
relaxation time \citep{Ch1942,SpitzerLyman1987}, 
\begin{equation}
t_{{\rm Ch}}\equiv\frac{\sigma^{2}}{\langle\left.(\Delta v_{\parallel})^{2}\rangle\right|_{v=\sqrt{3}\sigma}}=0.34\frac{\sigma^{3}}{G^{2}m^{2}n\ln\Lambda}\,,\label{eq:T-ch}
\end{equation}
where the second equality expresses the timescale for a typical test
star with velocity $v=\sqrt{3}\sigma$, to change its kinetic energy
by order unity, in a Maxwellian, isotropic and homogeneous stellar
system with stellar mass $m$, number density $n$ and velocity dispersion
$\sigma$. The Coulomb logarithm is defined by the ratio of maximal
and minimal impact parameters $\ln\Lambda=\ln(p_{\max}/p_{\min})$,
where $p_{\min}=Gm/\sigma^{2}$ (small angle approximation), and $p_{\max}$
is the size of the system.

These two timescales are related, but different, since $t_{{\rm Ch}}$
is local%
\footnote{The distinction between local and global is irrelevant for $t_{{\rm Ch}}$
when an infinite homogeneous system is assumed. %
} in physical space ($\mathbf{r},\mathbf{v}$), whereas $t_{E}$ is
local in $(E,J)$ space, and hence is orbit-averaged over all ($\mathbf{r},\mathbf{v}$)
values along the orbit, and often samples a wide range of conditions
in the stellar cluster. It is therefore to be expected that the values
of the two timescales can be quite different, even though they express
the same basic property of the system: systems with short relaxation
timescales evolve rapidly, and those with long timescales evolve slowly.
For example, when the two timescale are compared at $E\sim\sigma^{2}$,
typically $t_{\mathrm{Ch}}\gg t_{E}$ (Eq. \ref{eq:T-E-cusp}).

The diffusive timescales cannot be used in themselves to estimate
how long it takes an out-of-equilibrium system to approach steady
state, assuming that such exists. This generally depends on the specific
initial conditions, and the operative definition of convergence to
steady state. Numeric experiments show that in many cases $t_{\mathrm{Ch}}$,
evaluated at the radius of MBH influence, is a reasonable estimator
for the relaxation time to steady state, $t_{r}$. 

We now proceed to use the FP equation to demonstrate analytically
why that is so, and to derive the relation between $t_{E}$, $t_{\mathrm{Ch}}$
and $t_{r}$ in a power-law stellar cusp around a MBH\@. In Section
\ref{ss:longtimes} we verify that our analysis is not limited by
the approximate nature of the FP treatment, by numerically integrating
the system using MC simulations where the individual energy steps
are generated using the isotropic propagator $W_{E}(\DtE)$. Moreover,
we show that strong encounters, which are ignored in the FP treatment,
reduce the time to reach steady state by a factor of only $\lesssim4/3$.

Consider a background stellar population in steady state with a power-law
cusp $N_{bg}(E)=(5/4)N(E/E_{p})^{-9/4}$, where $N$ is the total
number of stars with $E>E_{h}$ . The DCs (Eqs.~\ref{eq:Gamma1-cusp-inf},~\ref{eq:Gamma2-cusp-inf})
are $\Gamma_{1}^{\Lambda}(E)=0$ and $\Gamma_{2}^{\Lambda}(E)\propto E^{9/4}$.
Assume a small initial perturbation superimposed at energy $E_{0}$
on the steady state background, $N_{p}(E)=N_{\star}\delta(E-E_{0})$.
This system describes, for example, test stars that are scattered
by a steady state distribution of field stars. Over time, the perturbation
will spread out until it relaxes to the $\propto E^{-9/4}$ steady
state distribution.

The dimensionless FP equation for the evolution of $N_{p}$ is 
\begin{eqnarray}
\frac{dN_{p}(x,\tau)}{d\tau} & = & \frac{1}{2}\frac{\partial^{2}}{\partial x^{2}}\{N_{p}(x,\tau)x^{9/4}\}\,,\label{eq:FP-prtrb}
\end{eqnarray}
where $x=E/E_{h}$ and $\tau=t/t_{E}(E_{h})$. We assume that the
number of stars with $E>E_{h}$ is constant (i.e.\  there is a reflective
boundary at $E_{h}$). The perturbation then evolves as (see Eq.~\ref{eq:N(x,t)-impulse-respond})
\begin{eqnarray}
\eta(x,\tau) & \equiv\frac{N_{p}(x,\tau)}{N_{p}^{\infty}(x)}\!-\!1 & =\sum_{n=1}^{\infty}\! Q_{n}(x)Q_{n}(x_{0})e^{-j_{5,n}^{2}\tau/128}\,,\label{eq:Time-dep-sol}
\end{eqnarray}
where $N_{p}^{\infty}(x)=N_{bg}(x)/N$, $Q_{n}(x)=\sqrt{x/5}J_{4}(j_{5,n}x^{-1/8})/J_{4}(j_{5,n})$,
$J_{n}(x)$ is the Bessel function of the first kind and $j_{n,m}$
is the $m$'th zero of $J_{n}(x)$. Similar solutions for a power-law
background cusps with $p>0$ are presented in Appendix \ref{s:FP-solution}.

The time it takes for the small perturbation to decay, and for the
system to converge back to its steady state, depends on the location
of the initial perturbation, and on the operative definition of the
convergence. The dynamical evolution back to steady state is slower
for higher $x_{0}$ (closer to the MBH), in spite of the fact that
the relaxation time decreases in that limit as $t_{E}(x)\sim x^{-p}$
($p=1/4)$. This counter-intuitive result can be understood by considering
a perturbation of mass $\Delta M$ at $x\ge x_{0}$, where the enclosed
steady state mass scales as $x_{0}^{p-3/2}$. It then follows the
fractional perturbation scales as $\delta M\propto x_{0}^{3/2-p}$,
and the rate for relaxing the perturbation scales as $\delta M/t_{E}\propto x_{0}^{3/2}$.
We therefore conservatively define the relaxation time as the time
it takes a $\delta$-function perturbation in the limit $x_{0}\to\infty$
to decay to an order unity perturbation, $\eta(x_{0,}t_{r})=1$, where
\begin{eqnarray}
\eta(t) & = & \lim_{x_{0}\to\infty}\sum_{n=1}^{\infty}Q_{n}^{2}(x_{0})e^{-j_{5,n}^{2}t/(128t_{E})}\,\nonumber \\
 & \approx & \frac{1}{1128\Gamma(5)^{2}}\frac{j_{5,1}^{8}}{J_{4}^{2}(j_{5,1})}e^{-j_{5,1}^{2}t/(128t_{E})}\,,\label{eq:eta}
\end{eqnarray}
where here $\Gamma$ is the Gamma function. This corresponds to the
relaxation time 
\begin{equation}
t_{r}=\frac{128}{j_{5,1}^{2}}\log\left(\frac{1}{1128\Gamma(5)^{2}}\frac{j_{5,1}^{8}}{J_{4}^{2}(j_{5,1})}\right)t_{E}\simeq11.1t_{E}\,.\label{eq:rlx-time}
\end{equation}

The conclusion that $t_{r}\sim O(10)t_{E}$ is robust, since it mostly
reflects the time it takes the perturbation to propagate down to $x=1$,
rather than the exact details of the convergence criterion. For example,
an alternative definition of $t_{r}$ in terms of evaporation, leads
to a similar result. Consider the evaporation of a perturbation through
the outer boundary $x=1$, and define the relaxation time to steady
state as the time it takes the number of stars still in the system
to decrease by a factor $P_{\star}=0.1$ (the conclusions depend only
logarithmically on the exact value of $P_{\star}$). In this setting
the probability to find the test star in the system is given by (see
Eq.~\ref{eq:N-p(X,t)})

\begin{equation}
\frac{N_{p}(x>1,\tau)}{N_{\star}}=\frac{2}{\sqrt{x_{0}}}\sum_{n=1}^{\infty}\frac{J_{5}(j_{4,n})J_{4}(j_{4,n}/x_{0}^{1/8})}{j_{4,n}J_{5}^{2}(j_{4,n})}e^{-j_{4,n}^{2}\tau/128}\,,\label{eq:N-absorb}
\end{equation}
 and the relaxation time (in the limit $x_{0}\to\infty)$ is given
by 
\begin{equation}
t_{r}\approx\frac{128}{j_{4,1}^{2}}\log\left(\frac{P_{\star}^{-1}j_{4,1}^{3}}{8J_{5}(j_{4,1})\Gamma(5)}\right)t_{E}\simeq10t_{E}\,,\label{eq:rlx-time-absorb}
\end{equation}
similar to the relaxation time we obtained before.

This simple model can also explain the behavior seen in a variety
of dynamical simulations of relaxation around a MBH (\citet{Bahcall1976}
Figure 2; \citet{Hopman2006} Figure 1; \citet{Madigan2011} Figure
30): the initial relaxation to steady state is extremely rapid (super-exponential),
while at later times it slows down to exponential decay. Eq. (\ref{eq:eta})
indeed shows that the perturbation converges to steady state as a
series of exponential decaying terms with progressively faster rates
$128/j_{5,n}^{2}=\{0.601,\,1.19,\,1.93,\,2.81,\,3.86,\,\dots\}$.
Thus, the higher terms vanish rapidly in a combined super-exponential
rate, leaving last only the first and slowest term, which decays exponentially
on a timescale $\gtrsim t_{E}$.

The relaxation time to steady state, $t_{r}$, can be related to the
Chandrasekhar time $t_{{\rm Ch}}$ in the host galaxy (idealized here
as an isotropic and homogeneous system). Beyond the MBH radius of
influence at $r_{h}=GM_{\bullet}/\sigma_{0}^{2}$, where $\sigma_{0}$
is the asymptotic velocity dispersion and the stars are unbound to
the MBH ($E<0$). The velocity distribution is Maxwellian, the DF
is given by Eq. (\ref{eq:boundary-out}), and therefore $t_{{\rm Ch}}$
is given by Eq. (\ref{eq:T-ch}), with the asymptotic stellar density
$n_{0}$. The typical orbital energy at $r_{h}$ is $\sim\sigma_{0}^{2}/2$.
Closer to the MBH, within the radius of influence where the dynamics
are Keplerian, the steady state DF for bound stars with $E>\sigma_{0}^{2}$
($\sim r_{h}/2$) is $f\approx2f\left(0\right)\left(E/\sigma_{0}\right)^{p}$
\citep[e.g.][]{Bahcall1976}. The energy diffusion timescale is then
(Eqs.~\ref{eq:tE-E},~\ref{eq:Gamma2-cusp-inf}) 
\begin{equation}
t_{E}(\sigma_{0}^{2})=\frac{1}{64}\frac{Q^{2}}{N\left(>\sigma_{0}^{2}\right)\log\Lambda}P\left(\sigma_{0}^{2}\right)\approx T_{{\rm Ch}}/44\,.\label{eq:T-E-cusp}
\end{equation}
It then follows that $t_{r}\sim10t_{E}\sim T_{{\rm Ch}}/4$ (Eqs.
\ref{eq:rlx-time}, \ref{eq:T-E-cusp}). The large numeric pre-factors
that relate these different diffusion timescales highlight the point
discussed above. While these timescales all generally reflect the
tendency of the system to evolve, they measure different aspects of
it: $t_{{\rm Ch}}$ is a \emph{local} quantity in physical space,
estimated \emph{outside} the evolving region; $t_{E}$ is an \emph{orbit-averaged}
quantity, estimated at some typical radius \emph{inside} the evolving
region; $t_{r}$ describes the timescale to reach \emph{global} steady
state. In many cases of interest, $t_{r}$ is not very different from
the age of the stellar system (or the Hubble time, a commonly assumed
upper limit). Inferences about the dynamical state of such systems
should be mindful of the large differences between the meanings and
values of the various diffusion timescales. The result $t_{r}\sim T_{{\rm Ch}}/4$
is consistent with numerical studies that show that a stellar system
around a MBH approaches steady state in about $T_{{\rm Ch}}/3$ to
$T_{{\rm Ch}}/2$ \citep[e.g.][]{Bahcall1976,Hopman2006}. 

The scaling of $t_{r}$ with MBH mass $M_{\bullet}$ can be obtained
from the relation $t_{r}\simeq10t_{E}$, the empirical $M/\sigma$
relation \citep{Ferrarese2000,Gebhardt2000,Tremaine2002}, 
\begin{equation}
M_{\bullet}=M_{0}\left(\sigma/\sigma_{0}\right)^{\alpha}\,,
\end{equation}
and the stellar number parametrization $N\left(>E\right)=Q\left(E/\xi\sigma^{2}\right)^{-5/4}$,
where $\xi$ is an order unity factor. The time to steady state is
then $t_{r}=\left(5\sqrt{2}\pi/64\right)\xi^{-5/4}\left(M_{0}/M_{\bullet}\right)^{3/\alpha}\left(GM_{\bullet}/\sigma_{0}^{3}\right)Q/\log Q$.
For $\alpha=4$, $M_{0}=1.3\times10^{8}M_{\odot}$, and $\sigma_{0}=200\,\mathrm{km}\mathrm{s}^{-1}$
\citep{Tremaine2002}

\begin{equation}
t_{r}=\left(7.8\times10^{9}\,\mathrm{yr}\right)\xi^{-5/4}\left(\frac{M_{\star}}{M_{\odot}}\right)^{1/4}\frac{\left(Q/10^{7}\right){}^{5/4}}{\left(\log Q/\log10^{7}\right)}\,.\label{eq:tr-M}
\end{equation}
This indicates that stellar cusps around lower-mass MBHs ($M_{\bullet}\lesssim10^{7}M_{\odot}$
with $M_{\star}=1\, M_{\odot}$), which have evolved passively over
a Hubble time, should be dynamically relaxed.

\section{Beyond the diffusion approximation}

\label{s:beyondFP}

\subsection{Kinematic limits on energy exchange}

\label{ss:physlim}

The transition probability encodes the full description of energy
evolution in the system due to 2-body encounters on all energy scales.
For hyperbolic Keplerian interactions between two stars, its general
form is $K_{r,E}\sim F_{r,E}(\Delta E)/|\Delta E|^{3}$, where $F_{r,E}$
is a non-diverging function of $\Delta E$, the energy change in an
encounter (see Appendix \ref{s:energy-transition}). The singularity
of $K_{r,E}$ at $\Delta E\rightarrow0$ is never reached in practice,
because the finite size of the system sets a lower limit on $\DE$.
In fact, an even stronger limit is implied by the restriction to hyperbolic
orbits. A test star on an orbit with energy $E$ has radial orbital
period $P(E)\sim GM_{\bullet}E^{-3/2}$ and a typical velocity $v\sim\sqrt{E}$.
An encounter with another star at impact parameter $b$ leads to an
exchange of energy of order $\Delta E\sim GM_{\star}/b$ and lasts
a time $\sim b/v$. Since the weakest encounter that can still be
approximated by a hyperbolic one, must have $b/v<P$, it follows that
$\DEmin\sim E/Q$ (i.e. $\Delta\epsilon_{\min}\sim1/Q$ for $\Delta\epsilon\equiv\Delta E/E$)
\citep{Bahcall1976}. This approximate effective low-$\Delta E$ cutoff
assumes that the contribution of distant and slow interactions is
negligible. It is estimated below empirically by $N$-body simulations
(see Section \ref{ss:Nbody}).

\begin{figure}
\noindent \centering{}\includegraphics[width=0.9\columnwidth]{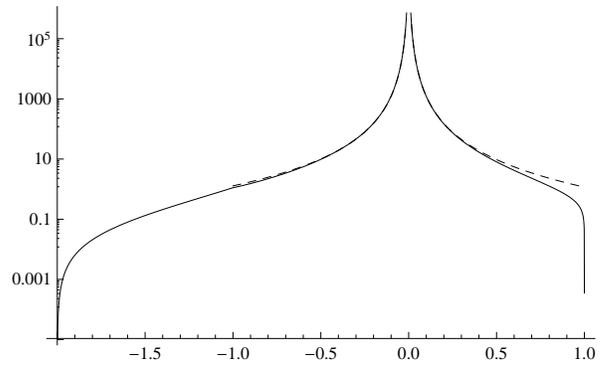}\caption{\label{f:Kfull}The transition probability $K_{r,E}$ (Eq. \ref{eq:K-rE})
for the case of $a/r=1$ and a BW cusp ($p=1/4$), as function of
$\Delta\epsilon$. The full transition probability (solid line) is
compared to the approximate form (Eq. \ref{eq:K-rE-exp}) (dashed
line). }
\end{figure}

Kinematics introduce an inherent high-$\Delta E$ limit \citep{GoodmanJ.1985}.
Strong encounters occur in the impulsive limit, where the distance
between the test star at $r$ and the background star at $r_{b}$
is very small. The energy is then exchanged over a short time and
the test star's position remains almost fixed. In that limit, the
energy one star can transfer to the other is at most its total kinetic
energy. Let the test star be on an initial Keplerian orbit with semi-major
axis $a$, energy $E=GM_{\bullet}/2a$ eccentricity $e$ and the eccentric
anomaly $\omega$, is at radius $r=a\left(1-e\cos\omega\right)$ at
the moment of the encounter. The most bound orbit it can be deflected
to at fixed $r$ has $a_{\min}^{\prime}=r/2$ and $E_{\max}^{\prime}>E$.
Thus, the maximal change in the relative energy of the test star is
$\Demax^{(+)}=\max\left|E^{\prime}-E\right|/E=\left(2a/r-1\right)$.
Conversely, the least bound orbit the test star can be deflected to
at fixed $r$ involves kicking a background star with an initial energy
$E_{b}$ to a maximally bound orbit with $a_{b}^{\prime}=r_{b}/2\approx r/2$.
Since typically the least bound (or unbound) background stars have
$E_{b}\sim0$, the maximal change in the relative energy in this case
is $\Demax^{(-)}\approx2a/r$. Figure (\ref{f:Kfull}) shows the full
transition probability for a specific value of $a/r$ and a power-law
DF; the kinematic high-$\Delta E$ limit is clearly seen. The high-$\Delta E$
limit is ultimately set by physical (inelastic) collisions between
stars of finite size, where orbital energy is not conserved due to
tidal interactions, mass loss, or stellar destruction. Orbit-averaging
over the phase of the test star yields $\left\langle \Demax^{(+)}\right\rangle =1$
and $\left\langle \Demax^{(-)}\right\rangle =2$, close to the commonly
assumed value $\Demax=1$. It is interesting to note that these averaged
kinematic limits are consistent with the small angle regime, conventionally
defined by an impact parameter $b>2GM_{\star}/\left\langle v\right\rangle ^{2}$
\citep[e.g.][]{BinneyJames2008}, which corresponds to $\Delta\epsilon<1$
.

Note however, that in the limit $e\rightarrow1$, $\Delta\epsilon$
diverges, and therefore the eccentricity-averaged transition probability
has finite contributions from arbitrarily large $\Delta\epsilon$.
In particular, the isotropic average transition probability, and the
resulting isotropic propagator, involve integrations over $\Delta\epsilon\rightarrow\infty$.

Note that the actual value of $\Demax$ may be lower than the kinematical
one, when destructive physical collisions between stars play a role.
In that case the maximal energy exchange is given by $\Demax^{\star}\sim4a/QR_{\star}\approx45\left(Q/4\times10^{6}\right)^{-1}\left(a/\mathrm{pc}\right)\left(R_{\star}/R_{\odot}\right)^{-1}$,
where $R_{\star}$ is the stellar radius. However, this is relevant
only extremely close to the MBH. For example, in the Galactic center
$\Demax<1$ for $a\lesssim22\,\mathrm{mpc}$, and even then the effect
of the collisional limit on the relaxation of the surviving stars
is small; for example at $a\sim0.22\,\mathrm{mpc}$, $\log\Lambda$
is reduced by only a factor of two.

\subsection{Relaxation on short timescales}

\label{sub:st-propagation}

In order to measure short-timescale evolution in $N$-body simulation,
it is necessary to derive the energy propagator in the short time
limit. Consider a test star with an initial energy $E$ and angular
momentum $J$, where the energy and angular momentum change as a result
of two-body interactions over a period of time $P(E)<t<t_{E}$. In
the small deflection angle limit (equivalently, $\Delta E/E<1$),
the orbit averaged energy transition probability, $K_{E,J}(\DE)=\int d\Delta JK_{E,J}(\DE,\Delta J)$,
can be approximated as (See Appendix~\ref{s:energy-transition} and~\citet{Goodman1983})
\begin{equation}
K_{E,J}(\DE)=\frac{\Gamma_{2}(E,J)+\Gamma_{1}(E,J)\DE+O(\DE^{2})}{2\left|\DE\right|^{3}}\,,\label{eq:K-JE}
\end{equation}
for $\DEmin<|\DE|<E$, where $\Gamma_{2}(E,J)$ and $\Gamma_{1}(E,J)$
are given by Eqs. (\ref{eq:Gamma1}) and (\ref{eq:Gamma2}). Note
that $\Gamma_{1,2}$ are shape parameters of the transition probability,
$K_{E,J}\left(\DE\right)$, and are not by themselves the DCs $\Gamma_{1,2}^{\Lambda}$,
which are related to them by 
\begin{equation}
\Gamma_{1,2}^{\Lambda}\left(E,J\right)=\Gamma_{1,2}(E,J)\log\Lambda\,.\label{eq:gamma-dc}
\end{equation}
This distinction is important because $\log\Lambda\sim\log Q$ is
typically a large, $O(10)$ factor, which is determined by our assumptions
that strong encounters are ignored. Later is section \ref{ss:longtimes}
we will address this assumption and estimate the contribution of strong
encounters. However we show below that the short timescales evolution
of the system does not depend on the high $\DE$-limit.

Assume that $t$ is short enough so that the background remains fixed,
and changes in $E$ and $J$ are small enough so that $K_{E,J}(\DE)$
is approximately constant along the trajectory of the test star in
($E$, $J$) space. Let $W_{E,J}(\DtE,t)$ be the energy propagator.
In that case the master equation is 
\begin{eqnarray}
\frac{\partial W_{E,J}(\DtE,t)}{\partial t} & = & \int K_{E,J}(\DE)W_{E,J}(\DtE-\DE,t)d\DE\nonumber \\
 &  & -q_{2b}(E,J)W_{E,J}(\DtE,t)\,,\label{eq:M-eq-short-time}
\end{eqnarray}
where $W_{E,J}(\DtE,0)=\delta(\DtE)$ and $q_{2b}$ is the total rate
at which 2-body encounters scatter the test star 
\begin{equation}
q_{2b}\left(E,J\right)=\int d\DE K_{E,J}(\DE)=\frac{1-\Lambda^{-2}}{2\DEmin^{2}}\Gamma_{2}(E,J)\,.\label{eq:te-1}
\end{equation}
The energy propagator $W_{E,J}(\DtE,t)$ can then be used to integrate
the master equation in a MC scheme \citep[e.g.][]{Shapiro1978}.

The Fourier transform of Eq.~\eqref{eq:M-eq-short-time} is 
\begin{equation}
\left(\frac{\partial}{\partial t}-\tilde{K}_{E,J}(k)+q_{2b}(E,J)\right)\tilde{W}_{E,J}(k,t)=0\,,\label{eq:Master-FT}
\end{equation}
where the solution is simply 
\begin{eqnarray}
W_{E,J}(\DtE,t) & = & \int_{-\infty}^{\infty}\frac{dk}{2\pi}e^{ik\DtE}e^{\left(\tilde{K}_{E,J}(k)-q_{2b}\right)t}\label{eq:W-FT}
\end{eqnarray}
and $\tilde{K}_{E,J}(k)$ is the Fourier transform of $K_{E,J}(\DE)$
\begin{eqnarray}
\tilde{K}_{J,E}(k) & = & q_{2b}-ik\Gamma_{1}^{\Lambda}-\frac{k^{2}}{2}\Gamma_{2}^{\Lambda}+\sum_{n=3}^{\infty}\frac{(-ik)^{n}}{n!}\Gamma_{n}^{\Lambda}\,,\label{eq:K-FT}
\end{eqnarray}
where $\Gamma_{n}^{\Lambda}$ are the $n$'th (non-central) moments
of $K_{J,E}(\DE)$. The higher odd and even moments are
\begin{equation}
\Gamma_{2n+1}^{\Lambda}=\frac{E^{2n}\Gamma_{1}}{2n}\frac{1-\Lambda^{-2n}}{\Demax^{2n}}\,,\,\,\,\,\,\Gamma_{2n+2}^{\Lambda}=\frac{E^{2n}\Gamma_{2}}{2n}\frac{1-\Lambda^{-2n}}{\Demax^{2n}}\,.\label{eq:DCsH}
\end{equation}

Note that Eq.~\eqref{eq:K-FT} has an analytical expression in terms
of cosine and sine integrals (see Eq.~\ref{eq:K-FT-App}). Thus a
full, non-perturbative, evaluation of the propagator $W_{E,J}(\DE,t)$
is obtained by numerically evaluating Eq.~\eqref{eq:W-FT}. This
is equivalent to including in the master equation DCs of \emph{all}
orders (i.e. all moments of the transition probability), albeit calculated
using an approximate transition probability, which incorporates the
large angle limit only as an effective cutoff. This is to be contrasted
with the standard FP treatment where only the first two DCs are included
(often also only as approximations).

Neglecting the contribution of the $n>2$ moments in Eq. \eqref{eq:K-FT}
results in the FP propagator

\begin{equation}
W_{E,J}^{FP}(\DtE,t)=\frac{1}{\sqrt{2\pi\Gamma_{2}^{\Lambda}t}}e^{-\frac{(\DtE-\Gamma_{1}^{\Lambda}t)^{2}}{2\Gamma_{2}^{\Lambda}t}}\,,\label{eq:W-FP}
\end{equation}
which can be related to the non-pertubative energy propagator (Eq.
\ref{eq:W-FT}) via Eq. \eqref{eq:K-FT},
\begin{eqnarray}
W_{E,J}(\DtE,t) & = & W_{E,J}^{FP}(\DtE,t)\left\{ 1+\right.\,\nonumber \\
 &  & -\frac{1}{2}\frac{1}{3!}\frac{1}{\log\Lambda}\frac{E\Gamma_{1}}{\Gamma_{2}}\sqrt{\frac{t_{E}}{t}}H\! e_{3}\left(\frac{\DtE-t\Gamma_{1}^{\Lambda}}{\sqrt{t\Gamma_{2}^{\Lambda}}}\right)\nonumber \\
 &  & +\frac{1}{2}\frac{1}{4!}\frac{1}{\log\Lambda}\frac{t_{E}}{t}H\! e_{4}\left(\frac{\DtE-t\Gamma_{1}^{\Lambda}}{\sqrt{t\Gamma_{2}^{\Lambda}}}\right)\nonumber \\
 &  & \left.+\dots\right\} \,,\label{eq:W-exp}
\end{eqnarray}
where $H\! e_{n}\left(x\right)$ are the Hermite polynomials and $t_{E}(E,J)=E^{2}/(\Gamma_{2}(E,J)\log\Lambda)$
is the energy diffusion time. Therefore, for $t>t_{E}/\log\Lambda$,
the higher moments can be neglected, as expected from the central
limit theorem. 

\begin{figure*}
\noindent \centering{}\includegraphics[width=0.9\columnwidth]{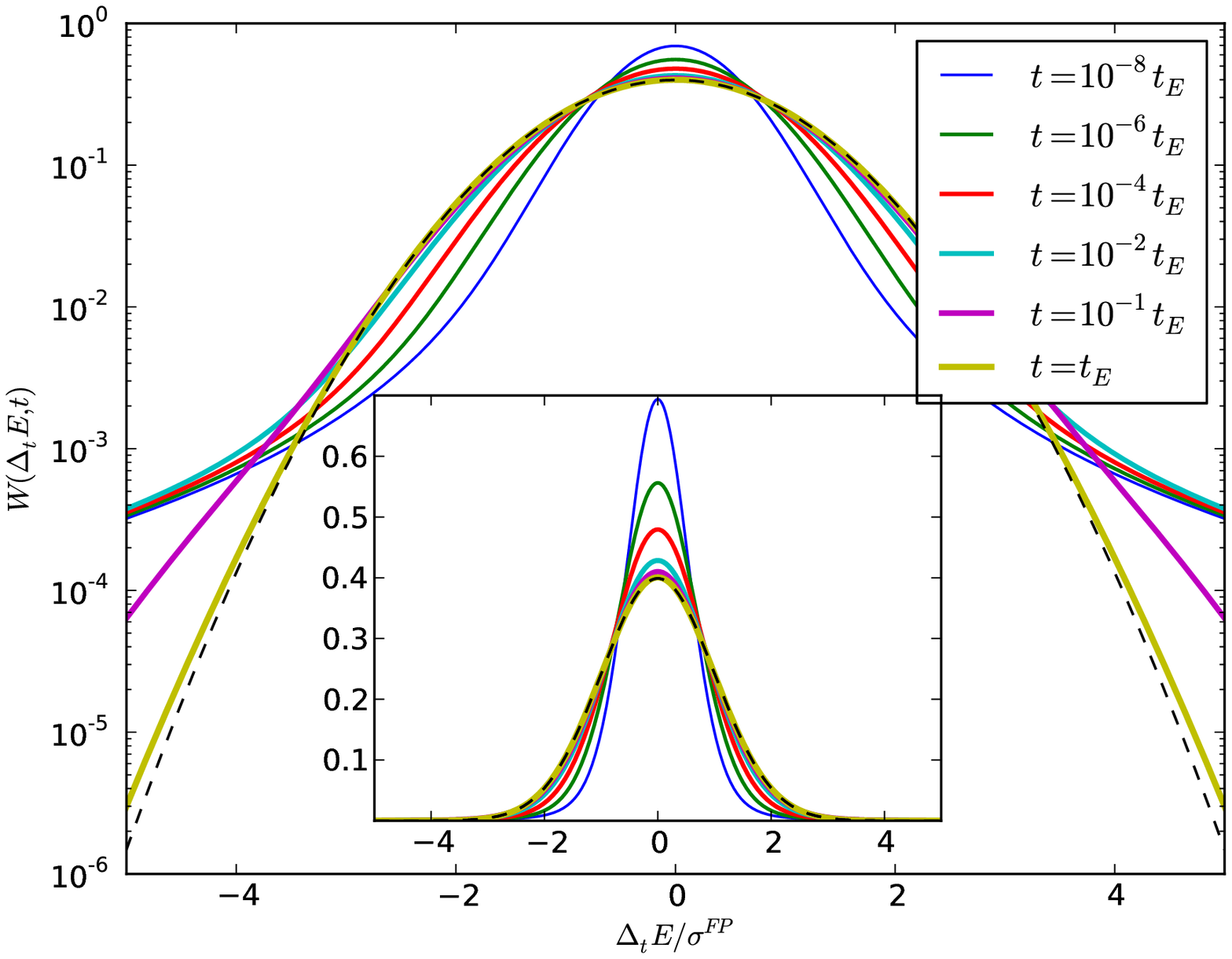}
\includegraphics[width=0.9\columnwidth]{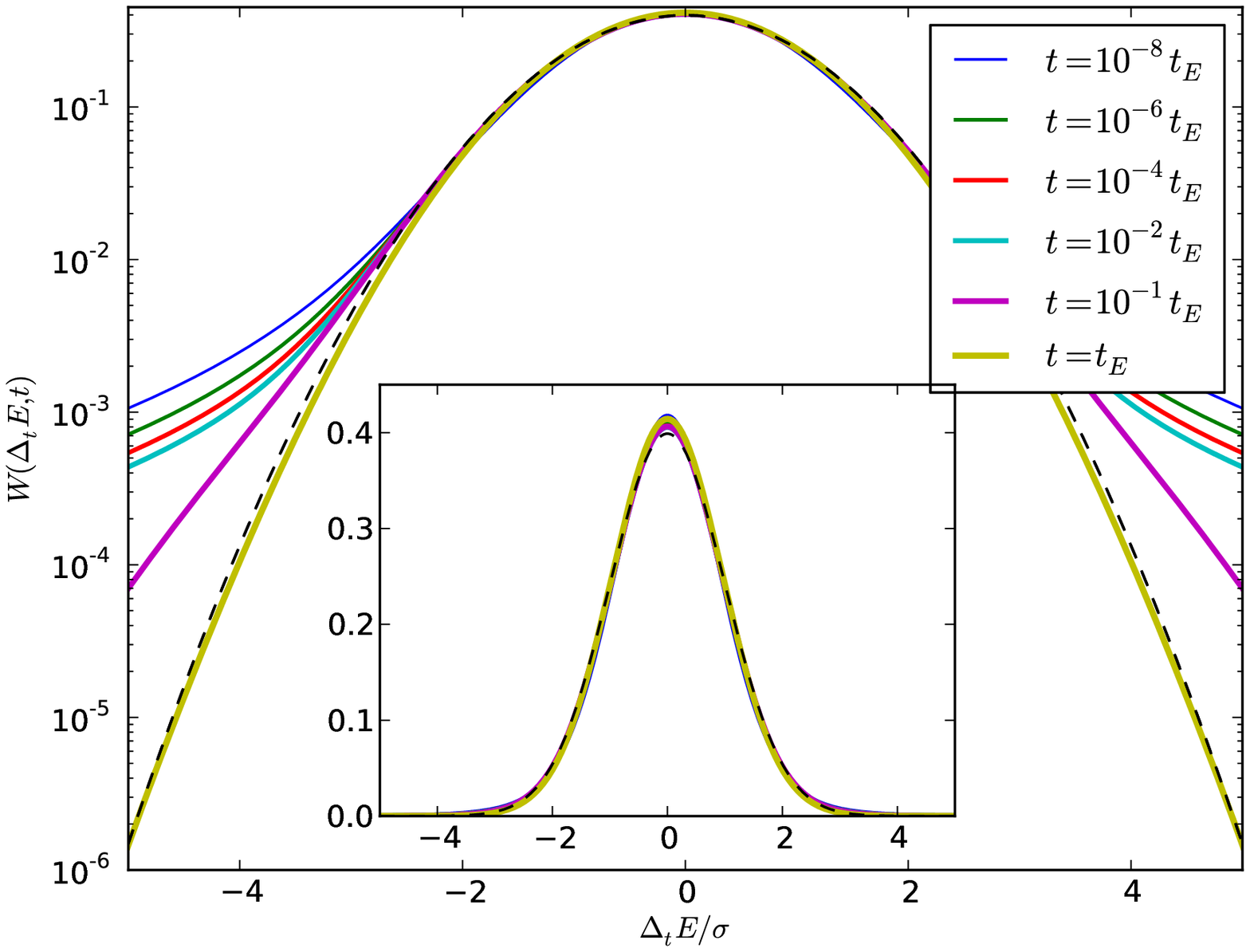} \caption{\label{fig:propagator-approx} Numerical evaluation of the propagator
(Eq.~\ref{eq:W-FT}) with $\DEmax=E$ and $\DEmin=E/10^{6}$ for
several time lags (solid lines). On the left the pdf is normalized
by $\sigma^{FP}=\sqrt{t\Gamma_{2}^{\Lambda}}=E\sqrt{t/t_{E}}$ as
expected from the FP approximation (Eq.~\ref{eq:W-FP}), on the right
the pdf is normalized by $\sigma=\sqrt{t\Gamma_{2}^{L}(t)}$ as expected
from the asymptotic analysis (first term in Eq.~\ref{eq:W-short-exp}).
A normalized Gaussian $\mathcal{N}(0,1)$ (dashed line) is shown for
comparison. The non-perturbative propagator converges to the approximate
FP one by $t\gtrsim0.1t_{E}$. The convergence is faster in the central
part of the distribution. The short timescale evolution of the central
part is well approximated by a Gaussian with $\sigma=\sqrt{t\Gamma_{2}^{L}(t)}$,
a clear demonstration of anomalous-diffusive evolution, which is faster
than the $\sigma\propto\sqrt{t}$ of a standard random walk process.}
\end{figure*}

\begin{figure}
\centering 

\includegraphics[width=1\columnwidth]{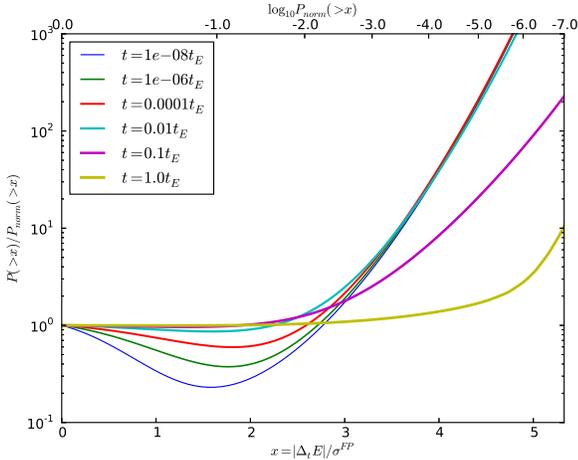}
 \caption{\label{fig:propagator-CDF-ratio} The probability $P(>x)$ for a $\left|\DtE\right|>x\sigma^{FP}$
event normalized by the probability to have such an event in a normal
distribution, where $\sigma^{FP}=E\sqrt{t\Gamma_{2}^{\Lambda}}$ defined
by the FP propagator. The FP propagator underestimates large $\sigma$
events, but over estimates the small $\sigma$ events.}
\end{figure}

On short timescales, the deviation from the FP propagator can be significant,
as seen in Figure~\ref{fig:propagator-approx}. Figure~\ref{fig:propagator-CDF-ratio}
shows the evolution of the energy propagator. On short timescales
($t\lesssim0.01t_{E}$), the FP approximation overestimates the contribution
of small energy changes and underestimates the contribution of large
energy changes to the accumulated change. Thus, formally rare events
(i.e. $>3\sigma$ for a Gaussian) have a much larger probability than
expected by the FP treatment. For example, the probability of a $5\sigma$
Gaussian event ($2.9\times10^{-7}$) can be actually as high as $1.6\times10^{-2}$.

It should be emphasized that the non-Gaussian, super-diffusive short-timescale
behavior occurs in spite of the fact that the physical energy transition
probability (in contrast to its simplified power-law approximation)
is in fact bounded by $\DEmax$, and therefore \emph{does have} non-diverging
moments. This seeming contradiction is reconciled by noting that on
short timescales, this cutoff is irrelevant, since it is too far away
in the wings of the distribution --- the system does not evolve enough
to ``be aware'' of its existence. This is a typical behavior for
distributions whose variance would diverge were it not for physical
cutoffs, for example truncated Levy-flights \citep[e.g.][]{Zumofen1994}.
Only on long enough timescales, when repeated scattering populates
the wings all the way to the cutoffs, does the variance converge,
the conditions for the application of the central limit theorem are
satisfied, and the propagator converges to its asymptotic $\sigma\propto\sqrt{t}$
behavior, independently of the functional form of $K_{E,J}(\DE)$.

In Appendix~\ref{s:propagator} we derive the effective untruncated
($\Demax\rightarrow\infty$) propagator $W_{E,J}^{s}(\DtE,t)$. Since
it includes arbitrarily strong energy exchanges, we refer to it at
the ``strong propagator'' (SP). This propagator is given by a heavy
tailed probability distribution 
\begin{eqnarray}
W_{E,J}^{s}(\DtE,t) & = & W_{E,J}^{0}(\DtE,t)+\frac{W_{E,J}^{1}(\DtE,t)}{4\log L\left(q_{2b}t\right)}\,,\label{eq:W-short-exp}
\end{eqnarray}
whose leading term is a Gaussian 
\begin{equation}
W_{E,J}^{0}(\DtE,t)=\frac{1}{\sqrt{2\pi t\Gamma_{2}^{L}(t)}}e^{-\frac{\left(\DtE-t\Gamma_{1}^{L}(t)\right)^{2}}{2t\Gamma_{2}^{L}(t)}}\,,\label{eq:W0}
\end{equation}
where 
\begin{equation}
\Gamma_{1,2}^{L}(t)=\Gamma_{1,2}(E,J)\log L\left(q_{2b}(E,J)t\right)\,\label{eq:Gamma12-tild}
\end{equation}
and $L\left(x\right)\approx\sqrt{e^{3-2\gamma_{E}}x}$, where $\gamma_{E}$
is the Euler constant. On long timescales approaching $t_{E}$, $L\left(q_{2b}t\right)\sim\Lambda$.
The second, correction term in \eqref{eq:W-short-exp} is the higher
order (in $1/\log L\left(q_{2b}t\right)$) even function that contributes
an asymptotic tail of $\sim|\DtE|^{-3}$ as $|\DtE|\to\infty$, 
\begin{eqnarray}
W_{E,J}^{1}(\DtE,t) & = & \frac{W_{c}^{1}\left(\frac{\DtE-t\Gamma_{1}^{L}(t)}{\sqrt{t\Gamma_{2}^{L}(t)}}\right)}{\sqrt{t\Gamma_{2}^{L}(t)}}\,,\label{eq:W1}
\end{eqnarray}
where $W_{c}^{1}(x)$ is defined in Eq. (\ref{eq:Wc1}). The next
order odd correction function corresponds to the drift term (see Eq.
\ref{eq:W-DE}) which is small on short timescales and is omitted
here. Unlike a Gaussian, which scales self-similarly as $\sigma(t)$,
the propagator in Eq. \eqref{eq:W-short-exp} is not self-similar
and thus does not have a single scaling. However, its central part
scales like $\sqrt{t\Gamma_{2}^{L}(t)}$, which increases with time
faster than $\sqrt{t\Gamma_{2}^{\Lambda}}$, that is, in a super-diffusive
manner. On longer timescales, when the cutoffs become important and
the limit $\Lambda\to\infty$ is not valid, Eq. (\ref{eq:W-short-exp})
is no longer a good approximation of the propagator. Note that on
all timescales this approximation is valid only for some central part
of the propagator and it is never a good approximation of the very
far tails. Therefore this approximated propagator can not be used
in the MC simulation even if the time steps of the MC are very short.

\subsection{Relaxation on long timescales}

\label{ss:longtimes}

The steady state configuration of a system, if such is attained on
cosmologically relevant timescales, has a special significance since
it is independent of the initial dynamical conditions, which are usually
unknown, and is typically the most likely configuration to be observed.
It is therefore important to understand how, and on what timescale,
a system reaches steady state, and how short-timescale evolution by
anomalous diffusion asymptotically approaches normal diffusion, as
described by the FP approximation.

\begin{table*}
\begin{centering}
\caption{\label{t:propagators}Summary of properties of isotropic propagators }

\par\end{centering}

\centering{}{\footnotesize }%
\begin{tabular}[t]{|l|c|c|c|c|}
\cline{2-5} 
\multicolumn{1}{l|}{\textbf{\footnotesize \rule{0em}{1.2em}}} & \textbf{\footnotesize Strong (SP)} & \textbf{\footnotesize Exact (EP)} & \textbf{\footnotesize Weak (WP)} & \textbf{\footnotesize Fokker-Planck (FP)}\tabularnewline
\hline 
\textbf{\footnotesize \rule{0em}{1.2em}Strong encounters} & {\footnotesize Overestimated} & {\footnotesize Exact} & \multicolumn{2}{c|}{{\footnotesize Not included}}\tabularnewline
\hline 
\textbf{\footnotesize \rule{0em}{1.2em}Relaxation time} & \multicolumn{1}{c}{{\footnotesize $t_{r}(\mathrm{SP})\,\,\,\,\,\le$}} & \multicolumn{1}{c}{{\footnotesize $t_{r}(\mathrm{EP})\,\,\,\,\,\le$}} & \multicolumn{1}{c}{{\footnotesize $t_{r}(\mathrm{WP})\,\,\,\,\,=$}} & {\footnotesize $t_{r}(\mathrm{FP})$}\tabularnewline
\hline 
\textbf{\footnotesize \rule{0em}{1.2em}Moments} & {\footnotesize All diverge} & {\footnotesize Only $n=1,2$ finite} & {\footnotesize All finite} & {\footnotesize $n=1,2$ finite, $n>2$ zero}\tabularnewline
\hline 
\textbf{\footnotesize \rule{0em}{1.2em}Gaussianity} & {\footnotesize Only central part, $\sigma^{2}\sim t\log t$} & \multicolumn{2}{c|}{{\footnotesize On long timescales, $\sigma^{2}\propto t$ }} & {\footnotesize Always, $\sigma^{2}\propto t$ $ $}\tabularnewline
\hline 
\end{tabular}
\end{table*}

\begin{figure}
\noindent \centering{}\includegraphics[width=1\columnwidth]{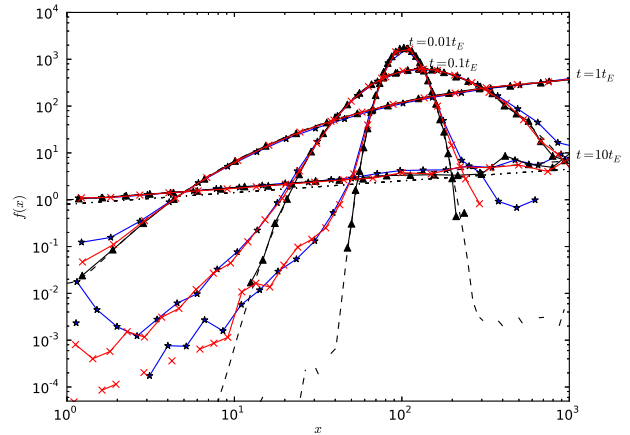}
\caption{\label{fig:fE-plot} Time evolution of the perturbation's DF $f_{p}(x,t)$
for initial perturbation at $x_{0}=100$. The system is integrated
by MC, with individual energy steps drawn from the FP propagator (triangles),
the isotropic WP $W_{E}(\DtE)$ (crosses), and the isotropic SP $W_{E}^{s}(\Delta_{t}E)$
(asterisks). The analytically derived DF $f_{p}(x,t)$ (Eq. \ref{eq:Time-dep-sol})
is shown for comparison (solid lines). All propagators converge to
the FP DF for $t>t_{E}$. The asymptotic steady state $f_{p}(x)\propto x^{1/4}$
is also shown for reference (dash-dotted line).}
\end{figure}

\begin{figure}
\noindent \centering{}\includegraphics[width=1\columnwidth]{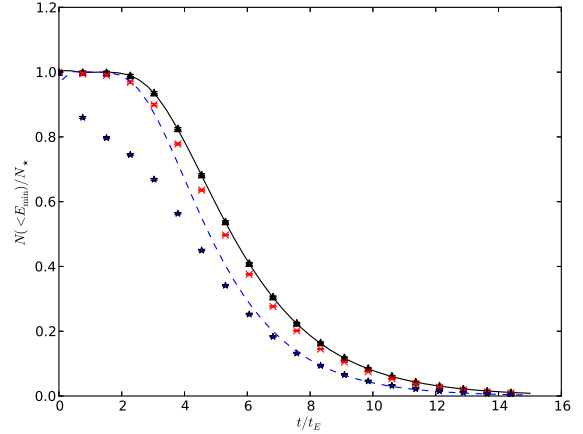}
\caption{\label{fig:N-tot} Time evolution of the perturbation's $N_{p}\left(x>1\right)/N_{\star}$
for initial perturbation at $x_{0}=10^{5}$ as a function of time.
The system is integrated by MC, with individual energy steps drawn
from the FP propagator (triangles), the isotropic WP $W_{E}(\DtE)$
(crosses), and the isotropic SP $W_{E}^{u}(\Delta_{t}E)$ (asterisks).
The analytically derived $N_{p}\left(x>1\right)/N_{\star}$ (Eq. \ref{eq:N-absorb})
is shown for comparison (solid line). The analytically derived $N_{p}\left(x>1\right)/N_{\star}$
(Eq. \ref{eq:N-absorb}) with $\Demax^{\mathrm{eff}}=e^{3/2-\gamma_{E}}\sqrt{10}$
is shown for comparison (dashed line). }
\end{figure}

On short timescales, rare strong encounters ($\Delta\epsilon>1$)
can be neglected. However, as time progresses, even these rare events
contribute significantly, and the kinematic limit (Section \ref{ss:physlim})
must be taken into account. A key issue is to estimate their effect
on the relaxation rate. In principle, the exact propagator (EP) provides
a full description of all encounters. However, from a practical point
of view it is difficult to implement it in an analytical or numerical
scheme, due to the required triple integration (but see \citealt{Goodman1983}).
Instead, as we show below, the EP can be bracketed by two limiting
approximations. The first, the SP $W_{E}^{s}(\Delta_{t}E)$ (Eq. \ref{eq:W-short-exp}),
over-estimates the contribution of strong encounters by taking the
weak-limit transition probability (Eq. \ref{eq:K-JE}), and integrating
it to $\Demax\rightarrow\infty$, i.e. beyond the weak limit (Eq.
\ref{eq:K-FT-App}). The second, the weak propagator (WP) $W_{E}(\Delta_{t}E)$,
integrates only up to $\Demax=1$, where the weak limit is valid (Eq.
\ref{eq:K-FT}). As argued below (see Figures \ref{fig:fE-plot},
\ref{fig:N-tot}), the SP, WP and FP propagator all lead to evolution
to the same steady state, on similar timescales. This then proves
that so must the EP, and that the contribution of strong encounters
to reaching steady state is not crucial. Table \ref{t:propagators}
summarizes the properties of the four isotropic propagators that enter
our analysis of relaxation.

The inclusion of strong encounters in the transition probability can
only accelerate relaxation. Therefore, the SP is expected to lead
to the fastest relaxation, followed by the EP, while the WP and the
FP are slowest, since they assume the weak limit. The question of
the existence and values of the moments of the propagators reflect
the form assumed for the transition probability, and in turn are related
to the nature of the long-term evolution of the system. The weak limit
assumption introduces a cutoff on the energy exchange, which ensures
that all moments exist. In particular, the neglect of all $n>2$ moments
assumed for the FP propagator, implies that it must be a Gaussian.
Both the WP and the exact one are guaranteed by the Central Limit
Theorem to converge to a Gaussian. The WP, by construction ($\Demax=1$)
converges to the FP Gaussian. This is not the case for the EP, which
takes into account the actual kinematic limits. These can be shown
to correspond in the asymptotic (Gaussian) long timescale limit to
an effective $\Delta\epsilon_{\max}^{\mathrm{eff}}\gg1$. However,
as we demonstrate below (Eq. \ref{e:tru}), this limit is approached
only on timescales much longer than the time for the stellar system
to relax to steady state, and therefore has little effect on the actual
time to reach steady state. This even holds for the SP ($\Demax\to\infty$),
where all the moments diverge and the propagator never converges to
a Gaussian with width $\propto\sqrt{t}$. 

The long-term evolution of the system can be described by repeated
application of the propagator on short-enough timescales. Computationally,
this can be realized by a MC simulation, which follows the evolution
of the orbital energy of individual stars as it changes by small random
increments, drawn from the distribution function described by the
propagator. Figure \ref{fig:fE-plot} shows snapshots from a set of
such simulations. In each, a different propagator is used to follow
the evolution of an initial $\delta$-function distribution of test
stars around a MBH, assuming a relaxed stellar background, and a reflecting
boundary condition at the outer edge of the system. In all cases,
the perturbation relaxes to the steady-state BW solution after $t_{r}\sim10t_{E}$.
This provides numeric validation of the analytical treatment of this
process (Section \ref{s:relaxation-times}). As expected, the evolving
DFs resulting from the WP and SP are significantly broader at early
times than the FP one. This reflects their heavy-tailed probability
distributions (e.g. Figure \ref{fig:propagator-approx}). 

Another demonstration of the near equivalence of the different propagators
is shown in Figure \ref{fig:N-tot}. where the different propagators
are used to follow the evolution of an initial $\delta$-function
distribution, assuming an absorbing boundary condition at the outer
edge of the system. In this configuration, the time to relaxation
can be defined as the time it takes the system to evaporate some large
fraction of its stars. Since we are interested here mainly in comparing
the evolution due to the different propagators, our conclusions do
not depend on the exact fraction. For definitiveness, we choose $0.9$.
Both the FP and the WP drive evaporation on almost the same timescale,
while the SP drives it only slightly faster, by a factor of $\sim1.1$
(for $\Demin=Q^{-1}=10^{-6}$). 

This modest acceleration in the evaporation rate by the SP, despite
the inclusion of arbitrarily strong encounters, can be explained by
the fact that the system evolves to its steady state much faster than
the rate at which $\De>1$ events contribute to relaxation. This can
be shown in a quantitative way by defining an effective, time-dependent
diffusion timescale for the SP.

On long timescales, the SP is dominated by its central Gaussian, and
therefore its width, $\sigma^{2}=t\Gamma_{2}\log\left[L\left(q_{2b}t\right)\right]$
(Eq. \ref{eq:W-short-exp}), can be used to estimate an instantaneous
diffusion timescale, $t_{E}^{\mathrm{ins}}(t)=E_{\min}^{2}/\Gamma_{2}\log L(q_{2b}t)$.
Therefore, at time $t$, the evolution of the system up to this time
can be described by the FP equation with an effective (time-averaged)
diffusion time 
\begin{equation}
\frac{1}{t_{E}^{\mathrm{eff}}\left(t\right)}=\frac{1}{t}\int_{0}^{t}\frac{1}{t_{E}^{\mathrm{ins}}(t^{\prime})}dt^{\prime}
\end{equation}
The analytic solution of the FP equation (see Section \ref{s:relaxation-times}),
$t_{r}\approx10t_{E}(E_{\min})$ then suggests to associate the time
for relaxation to steady state by the SP with the solution of the
\emph{ansatz} $t_{r}\approx10t_{E}^{\mathrm{eff}}(t_{r})$,

\begin{equation}
t_{r}^{SP}\approx10E_{\min}^{2}\left/\left[\Gamma_{2}\left(E_{\min}\right)\log\left(\sqrt{10}e^{3/2-\gamma_{E}}/\Demin\right)\right]\right.\,,\label{e:tru}
\end{equation}
which corresponds to an effective $\Demax=\sqrt{10}e^{3/2-\gamma_{E}}\approx8$.
This implies that by time $t_{r}$, the SP can be represented by an
FP propagator with a Coulomb logarithm that is formally increased
to $\approx\log(8/\Delta\epsilon_{\min})$ (Figure \ref{fig:N-tot}),
thereby providing a lower limit on the time to relaxation due to the
inclusion of strong encounters, $t_{r}\gtrsim t_{r}^{FP}\log(1/\Demin)/\log(8/\Demin)$. 

This lower limit is to be compared to the time for relaxation that
can be formally defined for the EP through its $\Gamma_{2}^{\Lambda}$
diffusion coefficient, where for the $p=1/4$ cusp, $\Lambda\approx56/\Demin$
(see Eq. \ref{eq:DC2-iso-inf-eff}). Assuming that the EP has already
converged to the diffusive limit, this then implies that $t_{r}^{EP}/t_{r}^{SP}\approx\log(8/\Demin)/\log(56/\Demin)<1$.
This is in contradiction with $t_{r}^{SP}$ being the lower limit
on the time to relaxation, which means that the EP does \emph{not}
reach the diffusive limit by time $t_{r}$, and therefore cannot be
fully described by an FP equation. Nevertheless, the FP equation remains
a reasonable approximation for describing relaxation around a MBH,
since the discrepancy in timescales is not very large. For example,
for $\Demin=Q^{-1}$ (Section \ref{ss:physlim}), $0.76<t_{r}^{FP}/t_{r}^{SP}<0.92$
in the range $Q=10^{3}-10^{10}\, M_{\odot}$.

\section{Measuring energy relaxation}

\label{s:measurements}

\subsection{Overview of the best fit procedure}

\label{ss:best-fit} 

The propagator $W_{E,J}(\DtE,t)$ (Eq.~\ref{eq:W-FT}) is determined
by the shape of the stellar cusp through the known functions $\Gamma_{1}(E,J)$
and $\Gamma_{2}(E,J)$ (Eqs.~\ref{eq:Gamma1},~\ref{eq:Gamma2}),
and by the parameters $\Demin$ and $\Demax$. However, on the short
timescales accessible through our $N$-body simulations, the propagator
is fully determined by $\Gamma_{1}$, $\Gamma_{2}$ and the parameter
$\Demin$ only (see Eq.~\ref{eq:W-short-exp}). In practice, $\Gamma_{1}(E,J)$
has a vanishing contribution to energy relaxation on short timescales,
so that the propagator can be approximated as a functional of $\Gamma_{2}$
with one free parameter, $\Demin$. 

As shown below, careful analysis of the $N$-body results reveals
that the propagator model provides an excellent description of the
data (Figures~\ref{fig:fm-fit} and~\ref{fig:W-data}). We thereby
validate the functional form of $\Gamma_{2}$, and determine the value
of $\Demin$. In principle, if it were also possible to validate $\Gamma_{1}$
and determine $\Demax$, then the full shape and range of the transition
probability (Eq. \ref{eq:K-JE}) could be determined, and the DCs
$\Gamma_{1,2}^{\Lambda}=\Gamma_{1,2}(E,J)\log\Lambda$ could be calculated
reliably. As it is, we have to rely on the fact that both $\Gamma_{1}$
and $\Gamma_{2}$ were derived in the same theoretical framework,
and therefore assume that the excellent fit for $\Gamma_{2}$ also
implies a validation of $\Gamma_{1}$. Furthermore, since in the small
deflection angle limit $\Demax\sim O(1)$, setting $\Demax=1$ is
expected to have only a small effect on the numerical value of $\log\Lambda=\log\left(\Demax/\Demin\right)$.
We find that with this empirically-motivated calibration, the numerical
values of the DCs lie within $\lesssim20\%$ of their theoretically
predicted values.

\subsection{Analysis of the $N$-body results}

\label{ss:Nbody}

To test and calibrate the analytic formulation of energy relaxation,
we carried out a suite of Newtonian $N$-body simulations. Each simulation
consisted of $10^{4}$ stars of equal mass $M_{\star}$, and a central
massive object of mass $M_{\bullet}=10^{6}M_{\star}$. The initial
semi-major axes of the stars were drawn from a $\rho(a)da\propto a^{2-\gamma}da$
distribution for $\gamma=1.0$, $1.25$, $1.5$ and $1.75$, with
random phases and isotropic orientations. The initial eccentricities
were drawn from an isotropic $\rho(e)de=2ede$ distribution. In the
Keplerian limit, these models approximately correspond to a power-law
density cusp, $n(r)\propto r^{-\gamma}$. The simulations were carried
out with the parallel \texttt{Nbody6++} code (implementing Hermite
integration with hierarchical block time steps, the Ahmad-Cohen neighbor
scheme for efficient calculation of accelerations, and the Kustaanheimo-Stiefel
regularization method for dealing with close few-body sub-systems)
\citep{Spurzem2001,Khalisi2006}.

Measuring energy relaxation in short-timescale $N$-body simulations
is challenging, since the distribution of energy jumps is heavy-tailed,
but the large energy exchange cutoff is not reached on these timescales
(Section~\ref{sub:st-propagation}). Consequently, commonly used
measures of evolution, such as the rms of the energy change, are strongly
affected by rare strong exchanges and do not offer a robust characterization
of the evolving energy distribution. After some experimentation, we
adopted a statistically robust method, described below, whose key
idea is to estimate the width of the central Gaussian component of
the energy distribution as the limit of fractional moments.

Snapshots of the $N$-body simulation results are recorded at equally
spaced times $\{t_{i}=i\delta t\}_{i=0,1,2,\ldots}$. For each star
$n$ of the total $N$, the orbital energy at $t_{i}$, $E_{i}^{n}$,
is calculated and recorded. The stars are assigned to logarithmically-spaced
energy bins by their initial orbital energy. For each star, and for
each time-lag $\{\Delta t_{k}=k\delta t\}_{k=1,2,3,\ldots}$, the
relative energy changes are collected on \emph{non-overlapping, consecutive
intervals}: $\{\De_{k}^{n}\}=\{(E_{i+k}^{n}-E_{i}^{n})/E_{i}^{n}\}_{i=0,k,2k,3k,\ldots}$,
thereby avoiding the complications of statistical inter-dependencies
and correlations. The measured relative energy differences, $\{\De_{k}^{n}\}$,
are drawn from the heavy-tailed parent distribution $W_{E_{i},J_{i}}(\De_{k}^{n},\Delta t_{k})$,
whose characteristic parameters need to be estimated from the numerical
data. We use the fact that the central part of $W(\De,t)$ is Gaussian
to a good approximation with a scale parameter $\sigma_{\DE}^{2}(t)=t\Gamma_{2}^{L}(t)$
(Eq.~\ref{eq:W-short-exp}). Since most of the data lies within the
central Gaussian (Figure~\ref{fig:W-data}), its width can be robustly
measured. As shown in Appendix~\ref{s:propagator}, the scale parameter
can be robustly estimated by fractional moments $\delta<1$ \citep[e.g.][]{Zumofen1994},
which give a larger weight to the central part of the distribution
and a smaller one to the diverging tails, 
\begin{eqnarray}
\sigma_{\DE}^{2} & (t)= & \lim_{\delta\to0}\frac{\langle|E(0)-E(t)|^{\delta}\rangle^{2/\delta}}{2\pi^{-1/\delta}\Gamma\left(\frac{1+\delta}{2}\right)^{2/\delta}}\,,\label{eq:fm-expected}
\end{eqnarray}
where $\Gamma(x)$ is the Gamma function and we assumed that $W_{E_{i},J_{i}}(\De,t)$
is symmetric since in our case, $t\Gamma_{1}^{2}\ll\Gamma_{2}$ (negligible
drift). In particular, this relation also holds for a pure Gaussian
(for any $\delta$). We verified that the results converge rapidly
as $\delta\rightarrow0$; the results below are estimated with $\delta=10^{-3}$.

\begin{figure}
\centering \includegraphics[width=1.05\columnwidth]{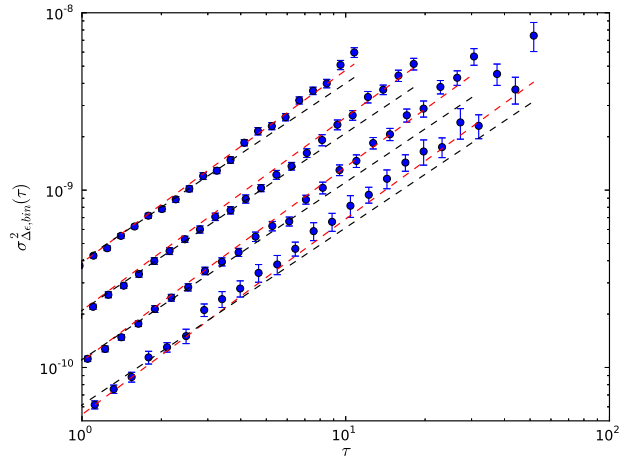}
\caption{\label{fig:correlation}The evolution of the energy scatter $\sigma_{\De,\mathrm{bin}}^{2}(\tau)$
(Eq. \ref{eq:sigma-bin}) (blue circles) in $N$-body simulations
for several energy bins (displayed shifted on the $y$-axis for convenience).
On time-lags $\tau>1$ the scatter grows as $\tau\log\tau$ (red dashed
lines), as expected from our analysis. This is somewhat faster than
the $\propto\tau$ rise expected for normal diffusion (black dashed
lines). }
\end{figure}

Over the short timescales of our simulations, $E$ and $J$ for each
star can be approximated as fixed at their initial values $E_{\star}$,
$J_{\star}$. Using Eq.~\ref{eq:fm-expected}, we define for each
star a dimensionless scale parameter, which we measure directly from
the data, 
\begin{equation}
\sigma_{\De,\star}^{2}=\frac{\Gamma_{2}^{L}(E_{\star},J_{\star},\tau P_{\star})\tau P_{\star}}{E_{\star}^{2}}=\lim_{\delta\to0}\frac{\langle|\De_{i,k}|^{\delta}\rangle^{2/\delta}}{2\pi^{-1/\delta}\Gamma\left(\frac{1+\delta}{2}\right)^{2/\delta}}\,,\label{eq:Gamma-star}
\end{equation}
where $\Delta t_{k}=\tau P_{\star}$, $P_{\star}=P(E_{\star})$ and
we use the minimal lag $\tau=1$ in order to obtain maximal statistics
(Figure \ref{fig:correlation}).

\begin{figure}
\centering \includegraphics[width=1\columnwidth]{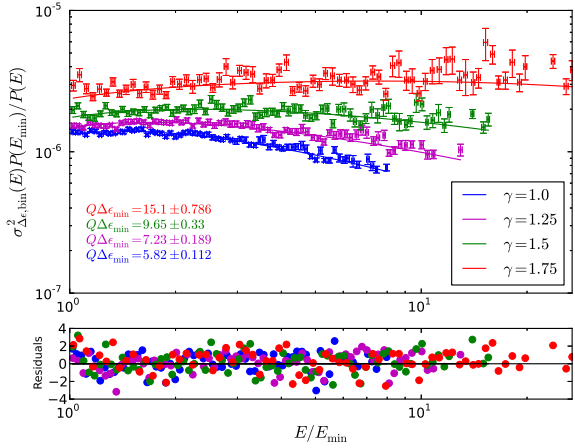}
\caption{\label{fig:fm-fit} The energy diffusion rate as measured by $\sigma_{\De,\mathrm{bin}}^{2}\left(\tau=1\right)P(E_{\min})/P\left(E_{\mathrm{bin}}\right)$
as a function of $E_{\mathrm{bin}}/E_{\min}$ for different cusps.
The analytical prediction (solid lines) is fitted to the measured
values in the $N$-body simulations (dots), using $\De_{\min}$ as
a free parameter (see Eq. \ref{eq:sigma-bin}).}
\end{figure}

We choose the energy bins to be wide enough so that the bin average
over $J$ is close enough to the isotropic average. Using Eq. (\ref{eq:Gamma-star}),
we define the bin averaged dimensionless scale parameter 
\begin{eqnarray}
\sigma_{\De,\mathrm{bin}}^{2}(\tau) & \equiv & \left\langle \frac{\Gamma_{2}^{L}(E_{\star},J_{\star},\tau P_{\star})\tau P_{\star}}{E_{\star}^{2}}\right\rangle _{_{bin}}\nonumber \\
 & \approx & \frac{\Gamma_{2}(E_{bin})\tau P_{\mathrm{bin}}}{E_{\mathrm{bin}}^{2}}\log L\left(\frac{\tau P_{\mathrm{bin}}\Gamma_{2}(E_{\mathrm{bin}})}{2\De_{\min}^{2}E_{\mathrm{bin}}^{2}}\right)\,,\label{eq:sigma-bin}
\end{eqnarray}
where $E_{bin}$ and $P_{bin}$ are the bin averaged values. The second
approximate equality relates the measured quantity to the theoretical
model (Eq. \ref{eq:Gamma12-tild}), under the assumption that the
weak dependence of $\log L(x)$ on $x$ allows $\left\langle x\log L(x)\right\rangle _{\mathrm{bin}}$
to be approximated by $\left\langle x\right\rangle _{\mathrm{bin}}\log L\left(\left\langle x\right\rangle _{\mathrm{bin}}\right)$
. Finally, we calculate $\Gamma_{2}$ from Eq. (\ref{eq:Gamma2-iso-cusp}),
and fit $\sigma_{\De,\mathrm{bin}}^{2}$ to the $N$-body data to
obtain the best fit estimate of $\Demin$ and a goodness of fit score
for the model (Figure \ref{fig:fm-fit}).

Our estimated best fit values for $\Demin$, averaged over $\sim10$
simulations each for various values of the power-law cusp, are presented
in Table~\ref{tbl:Lambda-Q}. We also list the relative errors resulting
from the simple approximation $\Lambda=Q$, assuming $\De_{\max}=1$.
Figure \ref{fig:correlation} shows the growth of the width of the
central Gaussian, $\sigma_{\De,\mathrm{bin}}^{2}$ as function of
the time-lag $\tau$. Pure diffusion would lead to a $\propto\tau$
behavior at $\tau>1$. However, the clear trend of a steeper than
linear growth $\sim\tau\log\tau$, a signature of anomalous diffusion.

\begin{table*}[tbh]
\caption{\label{tbl:Lambda-Q}Measured $\De_{\min}$}

\centering{}%
\begin{tabular}{l|c|c|c|c|c}
$\gamma$  & $\bar{\Delta\epsilon}_{\min}/Q^{-1}$ $^{(1)}$ & $\bar{\Lambda}/Q$ $^{(1,2)}$ & $\overline{\log\Lambda}/\log Q$$^{(1,2)}$  & $\overline{\chi_{\mathrm{red}}^{2}}(\bar{\Delta\epsilon}_{\min})$ & $n_{\mathrm{sim}}$ \tabularnewline
\hline 
$1.0$  & $5.3\pm0.5$  & $0.19\pm0.02$  & $0.88\pm0.01$  & $1.2\pm0.3$  & $8$ \tabularnewline
$1.25$  & $6.7\pm0.7$  & $0.15\pm0.02$  & $0.86\pm0.01$  & $1.3\pm0.2$  & $12$ \tabularnewline
$1.5$  & $9.2\pm1.3$  & $0.11\pm0.02$  & $0.84\pm0.01$  & $1.5\pm0.5$  & $9$ \tabularnewline
$1.75$  & $13.7\pm3.6$  & $0.07\pm0.02$  & $0.81\pm0.02$  & $2.0\pm0.6$  & $9$ \tabularnewline
\hline 
\multicolumn{6}{l}{{\scriptsize $^{1}$~For $Q=10^{6}$, best fit values for $N$-body
results (Section \ref{ss:Nbody}).}}\tabularnewline
\multicolumn{6}{l}{{\scriptsize $^{2}$~$\bar{\Lambda}=\Delta\epsilon_{\max}/\bar{\Delta\epsilon}_{\min}$,
where $\De_{\max}=1$}}\tabularnewline
\hline 
\end{tabular}
\end{table*}

Finally, to verify our analysis we plot the histogram of the orbital
energy jumps $\De_{i,k}$ in a given energy bin and compare it to
the bin-averaged propagator
\begin{eqnarray}
W_{E_{\mathrm{bin}}}(\De,t) & \simeq & \left\langle W_{E_{\star},J_{\star}}(\De,t)\right\rangle _{E_{\mathrm{bin}}},\label{eq:W-bin}
\end{eqnarray}
where we evaluated $W_{E_{\star},J_{\star}}(\De,t)$ numerically for
each star using the fitted $\DEmin$, and Eq.~\eqref{eq:Gamma2}.
An example is shown in Figure~\ref{fig:W-data}. For comparison we
also show the expected bin-averaged FP propagator $\langle W_{E_{\star},J_{\star}}^{FP}(\De,t)\rangle$
and the bin- averaged central Gaussian of the short timescale propagator
(Eq. \ref{eq:W-short-exp}).

\begin{figure}
\centering{} \includegraphics[width=1\columnwidth]{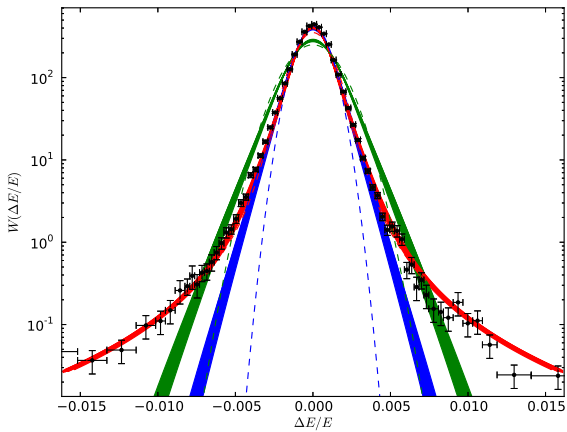}
\caption{\label{fig:W-data}The distribution of energy jumps (gray crosses)
measured in the $N$-body simulations, at one energy bin. Red strip:
exact numerical evaluation of the full propagator (Eq.~\ref{eq:W-FT}),
for the range of stellar angular momenta in the bin. Blue strip: the
central Gaussian of the full propagator (first order term in Eq.~\ref{eq:W-short-exp}).
Green strip: the FP propagator (Eq.~\ref{eq:W-FP}). }
\end{figure}

\section{Application example: Stellar orbits around the Galactic MBH}

\label{s:GCorbit}

The dynamical effects of stellar encounters between stars in the GC
may be detectable by next-generation telescopes \citep[e.g.][]{Weinberg2005,Bartko2009a},
and be used to probe the existence of the hypothesized stellar cusp
there. Since the relaxation time in the inner GC is many orders of
magnitude longer than the $O(10\,\mathrm{yr})$ observational timescale,
such encounters will be in the anomalous diffusion regime.

The question of the existence of the dark cusp near the MBH is closely
tied to the puzzling paucity of the old luminous giants there, whose
absence is in contradiction to the theoretical expectation for an
old relaxed system \citep{Bahcall1976,Bahcall1977,Alexander2009}.
A possible explanation is that a relaxed cusp does exist there, but
for some reason, for example the selective destruction of extended
giant envelopes \citep[e.g.][]{Alexander1999a}, the old giants do
not trace the overall population. In that case the cusp is dark, that
is, composed mainly of compact remnants and faint low-mass stars,
and is expected to be dominated by $\sim10\, M_{\odot}$ stellar mass
black holes that have accumulated by the process of strong mass segregation
over cosmological times \citep{Alexander2009}. 

Changes in orbital energy can be expressed as changes in the radial
orbital period (equivalently, a phase drift). Assuming that the dominant
cause of such changes is 2-body interactions, or that it is possible
to model and subtract all other competing effects, then one way of
detecting or constraining the dark cusp is through the stochastic
orbital phase drift in the few observed luminous stars very close
to the MBH.

\begin{figure}
\centering \includegraphics[width=0.96\columnwidth]{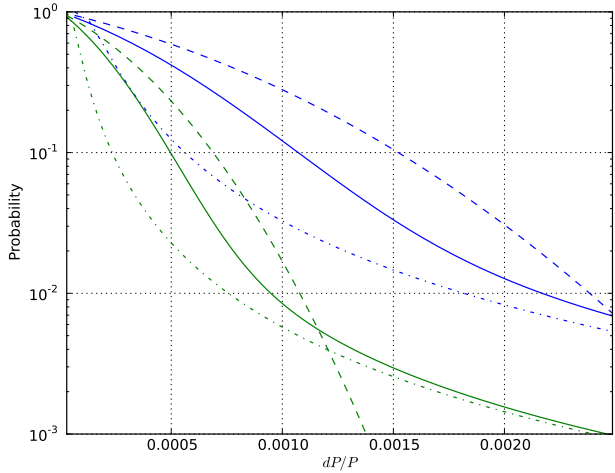}
\caption{\label{fig:dP-prob} Probability for a change in the orbital period
of a star with a $1$ year period, over an observation time of $10$
years, for several background population models. The probability for
an accumulated period change due to multiple energy changes calculated
by the short time propagator (Eq.~\ref{eq:W-FT}) (solid lines),
is larger than the probability for this change due to a single energy
scatter (dash-dotted lines). For comparison, we show the accumulated
period change probability calculated by the FP propagator (dashed
line). The background population is modeled by a cusp of $1M_{\odot}$
stars with a slope of $\gamma=1.75$ (green) with a total mass of
$10^{6}M_{\odot}$ within $1$ pc, and by a cusp of $10M_{\odot}$
stellar remnants with a slope of $\gamma=1.9$ (blue) and a total
mass of $10^{5}M_{\odot}$ within $1$ pc.}
\end{figure}

\begin{figure}
\centering \includegraphics[width=1\columnwidth]{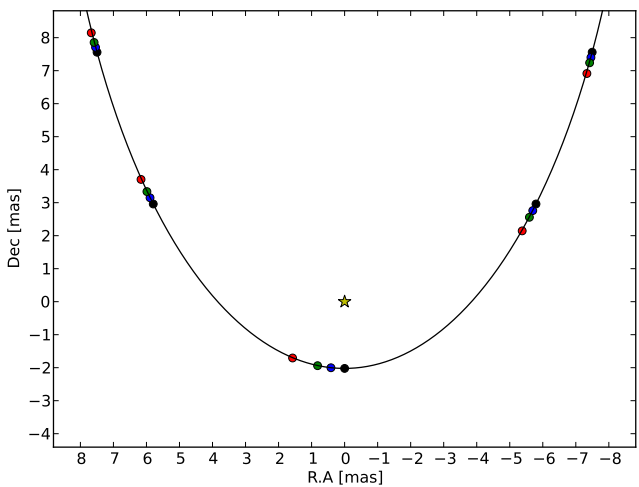}
\caption{\label{fig:prtb-orbit} Visualization of the orbital perturbation
of a one year Keplerian orbit with an eccentricity of $0.9$ to a
level of $\Delta P/P=0.0015$ (blue), $0.003$ (green) and $0.006$
(red), sampled on equal times, with respect to the original period
(black). The maximal spatial shifts are $0.415$ mas, $0.824$ mas
and $1.61$ mas.}
\end{figure}

Figure~\ref{fig:dP-prob} shows the probability of a period change
$dP/P$ after $10$ years for a test star with an orbital period of
one year (semi-major axis of $0.7$ mpc), as a result of interacting
with a relaxed main sequence cusp ($\rho(r)\sim r^{-7/4}$) and with
a mass-segregated dark cusp ($\rho(r)\sim r^{3/2+p}$) of stellar
remnants. Figure~\ref{fig:prtb-orbit} shows that a change of $dP/P=0.0015$
over 10 years due to a mass segregated dark cusp, measurable by next-generation
telescopes, has a non-negligible probability of $\approx3.3\%$ per
star. It should be noted that assuming only the effects of strong
single encounters leads to an \emph{under-estimate} of the probability
(1.5\%) of such an event, while the FP probability of $10.5\%$ is
an \emph{over-estimate}, because the event lies at the relatively
under-populated intermediate wings of the propagator (Figure ~\ref{fig:dP-prob}).
The naive FP estimate could therefore lead to erroneous conclusions
about the absence of a cusp in the case of a null result. For example,
if an event of $dP/P=0.0015$ is detected, the FP $\left\{ 1,2,3,4,5\right\} \sigma$
confidence levels on the minimal number of stars in the cusp, $\min_{n\sigma}N_{FP}$,
will differ from the true ones (due to anomalous diffusion), $\min_{n\sigma}N$,
by factors $\min_{n\sigma}N_{FP}/\min_{n\sigma}N=\left\{ 0.52,0.55,1.7,38,2.7\times10^{3}\right\} $,
respectively. Thus, the erroneous assumption of normal diffusion may
result in either over- or under-estimates of the dark cusp density,
depending on the chosen confidence level.


\section{Discussion and summary}

\label{s:summary}

The stochastic evolution of a stellar system is described at the most
basic level by the transition probability, which expresses all the
physical processes considered, and all the approximations assumed.
The accumulated effect of multiple scattering events over a time that
is short enough for the system to remain almost unchanged, is constructed
by integration over the transition probability, and is described by
the propagator. This general description of the short- or medium-term
evolution can lead to a wide range of stochastic processes. Only in
the specific case when the propagator asymptotes to a Gaussian, which
is fully defined by its two first moments (the first and second DCs),
does the evolution progress as $\propto\sqrt{t}$ and the process
becomes normal diffusion, which is described by the FP equation. However,
it is not at all guaranteed that this convenient and useful limit
provides a valid description of the accumulated effects of the transition
probability---this has to be proved case by case. There are in fact
many examples of anomalous diffusive processes, whose origins can
be traced to divergences of the moments of the transition probability,
such as the super-diffusive Levy flights \citep[e.g.][]{Metzler2000}

The FP equation has been applied extensively in past studies of nuclear
dynamics near a MBH. However, the simplifications, assumptions and
free parameters (many necessitated by the diverging, un-screened nature
of the gravitational force), which are needed to justify and use this
method, remained largely untested. This called for a more rigorous
validation of the basic underlying assumptions. This need was further
motivated by some conflicting theoretical and numerical results on
the emerging timescales of diffusion and relaxation in galactic nuclei
\citep[e.g.][]{Bahcall1976,PretoMiguel2004,Merritt2010b,Madigan2011}.
These impact multiple issues of interest: the rates of tidal disruptions,
the rates of GW emission events from inspiralling compact objects,
and the stellar distributions very close to MBHs, and in particular
those of compact remnants and stars on relativistic orbits. Furthermore,
the one system where the stellar distribution around an MBH can be
directly observed, the GC, does not follow the FP predictions for
an isolated old population, in contrast to prior expectations \citep{Buchholz2009,Do2009,Bartko2010,Merritt2010b}. 

Past studies of relaxation theory usually bypassed the use of the
transition probability \citep[but see][]{Goodman1983}. They assumed
from the outset that the diffusion coefficients exist and only the
first two are important, and calculated them by adopting cutoffs (the
small angle and hyperbolic orbits approximations) for the exchanged
energy in 2-body interactions. However, this procedure limits the
possibility of describing the behavior on short-timescales, of verifying
that relaxation does in fact converge to normal diffusion, and of
estimating the contribution of strong encounters.

We focused here on the short-timescale behavior of energy relaxation,
which can be directly probed by high-accuracy $N$-body simulations,
and is easier to describe analytically in terms of small deviations
from an almost stationary DF. We derived a rigorous analytical description
of relaxation on short-timescales, which we validated by fitting to
numerical results using a robust statistical analysis method. This
then made it possible to tie the short-timescale evolution, calibrated
by the $N$-body experiments, to the asymptotic FP one, thereby fixing
the absolute timescales for diffusion and relaxation around a MBH
to within $O(10\%)$. It should be emphasized that the convergence
of the propagator to the central limit, and the return of the stellar
system to steady state, are in principle unrelated processes, which
proceed at unrelated rates. There is therefore no a-priori guarantee
that the FP treatment is a good approximation for describing relaxation.
However, we showed that in the weak limit, the evolution of the propagator
converges to the FP description long before the system relaxes to
its steady state. We also showed that the inclusion of strong collisions
shortens the convergence time by less than a factor of $4/3$, because
the system reaches steady state before these rare collisions have
time to play a substantial role. Together these results validate the
use of the FP approximation on the intermediate and long timescales,
and prove that a perturbed BW cusp will regain is steady state after
$t_{r}\sim10t_{E}$, where $t_{E}$ is the diffusion time at the outer
boundary. We conclude that stellar cusps around lower-mass MBHs ($M_{\bullet}\lesssim10^{7}M_{\odot}$),
which have evolved passively over a Hubble time, should be dynamically
relaxed. Anomalous diffusion can affect processes observed on short
timescales. As an example, we showed that if future observation of
stars orbiting the Galactic MBH detect orbital energy perturbations
due to the dark cusp of stellar remnants that is expected to be there,
estimates of the cusp density will depend critically on accounting
for the anomalous nature of the diffusion.

Several open question remain. In this study we restricted ourselves
to energy relaxation in a single mass population. A mass spectrum
will generally change the steady state and shorten the relaxation
time \citep[e.g.][]{Alexander2009,Amaro-Seoane2010a}. It is likely
that the convergence of the propagators will depend on mass, and the
use of the multi-mass FP equation will then have to be re-examined
and justified. The evolution of angular momentum, which is crucial
for replenishing stars destroyed by the MBH, and thus competes with
energy relaxation, was not addressed here at all. The same considerations
of anomalous diffusion should in principle apply also to angular momentum.
This requires further study.

\section{Acknowledgments}

We are grateful to Rainer Spurzem for the use of the \texttt{Nbody6++}
code. T.A. acknowledges support by ERC Starting Grant No. 202996 and
DIP-BMBF Grant No. 71--0460--0101.

\appendix

\section{The energy transition probability}

\label{s:energy-transition}

We derive here explicit expressions for the energy transition probability
for several specific cases. The energy transition probability is generally
given by \citet{Goodman1983} Eq. A11 %
\footnote{We correct here a couple of printing errors in Eq. A11 of \citet{Goodman1983}.
The correct prefactor should have a $\pi^{2}$ term, and not $\pi$
(cf Eq. A7 there). The first term of the second line should have $v^{\prime}$
terms and not $v$ (cf Eq. A12 there).%
} 
\begin{eqnarray}
K_{r}\left(E,E^{\prime}\right) & = & \frac{2\pi^{2}G^{2}m^{2}}{v\left|\DE\right|^{3}}\begin{cases}
\int_{E_{\min}}^{E^{\prime}}dE_{f}f\left(E_{f}\right)\left(\frac{8}{3}v^{2}-4\Delta E\right)v+\int_{E+\DE}^{\varphi\left(r\right)+\DE}dE_{f}f\left(E_{f}\right)\left(\frac{8}{3}v_{f}^{\prime2}-4\Delta E\right)v_{f}^{\prime} & \DE<0\\
\int_{E_{\min}}^{E^{\prime}}dE_{f}f\left(E_{f}\right)\left(\frac{8}{3}v^{\prime2}+4\DE\right)v^{\prime}+\int_{E+\Delta E}^{\varphi\left(r\right)}dE_{f}f\left(E_{f}\right)\left(\frac{8}{3}v_{f}^{2}+4\Delta E\right)v_{f} & \DE>0
\end{cases}\ ,\label{eq:K-r}
\end{eqnarray}
where $E^{\prime}=E+\DE$, $v_{f}=\sqrt{2(\phi-E_{f})}$, $v_{f}^{\prime}=\sqrt{2(\phi-E_{f}+\DE)}$,
$v=\sqrt{2(\phi-E)}$ and $v^{\prime}=\sqrt{2(\phi-E^{\prime})}$.
Note that here $E$ is the positively defined orbital energy and $\phi\left(r\right)$
is positively defined potential. Thus Eq. \eqref{eq:K-r} reads 
\begin{eqnarray}
K_{r,E}(\DE) & = & \frac{32}{3}\frac{\pi^{2}G^{2}m^{2}}{\left|\Delta E\right|^{3}}\left(\varphi-E\right)\nonumber \\
 &  & \begin{cases}
\left(1-\frac{3}{4}\frac{\Delta E}{\varphi-E}\right)\int_{E_{\min}}^{E+\Delta E}dE_{f}f\left(E_{f}\right)\\
+\int_{E+\DE}^{\varphi\left(r\right)+\DE}dE_{f}f\left(E_{f}\right)\left(\frac{\varphi-E_{f}}{\varphi-E}\right)^{3/2}\left(1+\frac{1}{4}\frac{\DE}{\varphi-E_{f}}\right)\sqrt{1+\frac{\DE}{\varphi-E_{f}}} & \DE<0\\
\left(1-\frac{1}{4}\frac{\Delta E}{\varphi-E}\right)\sqrt{1-\frac{\DE}{\varphi-E}}\int_{E_{\min}}^{E+\DE}dE_{f}f\left(E_{f}\right)\\
+\int_{E+\DE}^{\varphi\left(r\right)}dE_{f}f\left(E_{f}\right)\left(\frac{\varphi-E_{f}}{\varphi-E}\right)^{3/2}\left(1+\frac{3}{4}\frac{\DE}{\varphi-E_{f}}\right) & \DE>0
\end{cases}\,.\label{eq:K-rE}
\end{eqnarray}
Assuming $|\DE|<E$ and expanding Eq. (\ref{eq:K-rE}) in $\DE$,
we obtain 
\begin{eqnarray}
K_{r,E}(\DE) & = & \frac{32}{3}\frac{\pi^{2}G^{2}m^{2}}{|\DE|^{3}}\times\nonumber \\
 &  & \left\{ (\varphi-E)\left(\int_{E_{\min}}^{E}dE_{f}f(E_{f})+\int_{E}^{\varphi(r)}dE_{f}f(E_{f})\left(\frac{\varphi-E_{f}}{\varphi-E}\right)^{3/2}\right.\right)\nonumber \\
 &  & -\frac{3}{4}\left(\int_{E_{\min}}^{E}dE_{f}f(E_{f})-\int_{E}^{\varphi(r)}dE_{f}f(E_{f})\sqrt{\frac{\varphi-E_{f}}{\varphi-E}}\right)\DE\nonumber \\
 &  & \left.-\frac{3}{4}f(E)\DE^{2}+O\left((\DE)^{3}\right)\right\} \,.\label{eq:K-rE-exp}
\end{eqnarray}

Taking the orbit average $K_{J,E}(\DE)=2P^{-1}\int_{rp}^{ra}K_{r,E}(\DE)dr/v_{r}$,
we obtain 
\begin{equation}
K_{E,J}\left(\DE\right)=\frac{1}{2}\frac{1}{|\DE|^{3}}\left(\Gamma_{2}(E,J)+\Gamma_{1}(E,J)\DE+\Gamma_{0}(E)\DE^{2}+O\left((\DE)^{3}\right)\right)\,,\label{eq:K-JE-xpn}
\end{equation}
where 
\begin{equation}
\Gamma_{0}(E)=-16\pi^{2}G^{2}m^{2}f(E)\,,\label{eq:Gamma0}
\end{equation}
 
\begin{equation}
\Gamma_{1}(E,J)=\frac{32\pi^{2}G^{2}m^{2}}{P}\int_{r_{p}}^{r_{a}}\frac{dr}{v_{r}}\left[\int_{E}^{\varphi\left(r\right)}dE_{f}f\left(E_{f}\right)\left(\frac{\varphi-E_{f}}{\varphi-E}\right)^{1/2}-\int_{E_{\min}}^{E}dE_{f}f\left(E_{f}\right)\right]\,,\label{eq:Gamma1}
\end{equation}
and 
\begin{equation}
\Gamma_{2}(E,J)=\frac{128\pi^{2}G^{2}m^{2}}{3P}\int_{r_{p}}^{r_{a}}\frac{dr}{v_{r}}\left(\varphi-E\right)\left[\int_{E}^{\varphi\left(r\right)}dE_{f}f\left(E_{f}\right)\left(\frac{\varphi-E_{f}}{\varphi-E}\right)^{3/2}+\int_{E_{\min}}^{E}dE_{f}f\left(E_{f}\right)\right]\,.\label{eq:Gamma2}
\end{equation}

Assuming that the DF is isotropic in velocity space, by averaging
over angular momentum 
\begin{equation}
\langle X\rangle_{iso}=\frac{1}{J_{c}^{2}}\int_{0}^{J_{c}^{2}}XdJ^{2}=\frac{2}{J_{c}^{2}}\int_{0}^{v}Xr^{2}v_{r}dv_{r}\,,\label{eq:J-avg}
\end{equation}
we obtain 
\begin{eqnarray}
\Gamma_{1}(E) & = & -16\pi^{2}G^{2}m^{2}\bar{f}(E)E\left[\int_{-\infty}^{1}g(s)ds-\int_{1}^{\infty}\frac{g(s)}{s^{5/2}}ds\right]\,,\label{eq:Gamma1-iso}
\end{eqnarray}
 
\begin{eqnarray}
\Gamma_{2}(E) & = & \frac{64\pi^{2}G^{2}m^{2}}{3}\bar{f}(E)E^{2}\left[\int_{-\infty}^{1}g(s)ds+\int_{1}^{\infty}\frac{g(s)}{{s}^{3/2}}ds\right]\,,\label{eq:Gamma2-iso}
\end{eqnarray}
where $g(s)=\bar{f}(sE)/\bar{f}(E)$.

An isotropic averaged transition probability can be directly obtained
from Eq. \eqref{eq:K-rE} without assuming the weak limit, 
\begin{equation}
K_{E}\left(\Delta E\right)=\frac{4}{PJ_{c}^{2}}\int_{0}^{v}dv_{r}\int_{r_{p}}^{r_{a}}\frac{dr}{v_{r}}r^{2}v_{r}K_{r,E}(\DE)=\frac{\mathcal{K}_{E}}{\left|\DE\right|^{3}}H\left(\De\right)\,,
\end{equation}
where $\mathcal{K}_{E}=\frac{32\pi}{3}G^{2}m^{2}\bar{f}\left(E\right)E^{2}$
and

\begin{eqnarray}
H\left(\De\right) & = & \begin{cases}
\left(1-\frac{3}{4}\Delta\epsilon\right)\int_{-\infty}^{1}g\left(s+\De\right)ds+\int_{1}^{\infty}\left(s-\frac{3}{4}\De\right)\frac{ds}{s^{5/2}}g\left(s+\De\right) & \De<0\\
\frac{\left(1+\frac{7}{4}\De\right)}{\left(1+\De\right)^{5/2}}\int_{-\infty}^{1}g\left(s+\De\right)ds+\int_{1}^{\infty}\left(s+\frac{7}{4}\De\right)\frac{ds}{\left(s+\Delta\epsilon\right)^{5/2}}g\left(s+\De\right) & \De>0
\end{cases}\,.\label{eq:H-de}
\end{eqnarray}
Since isotropic averaging includes $J\to0$ where $\De$ diverges,
the DCs are given by
\begin{eqnarray}
\Gamma_{1}^{\Lambda}\left(E\right) & = & \mathcal{K}_{E}\left(E\right)\left(\int_{\Demin}^{\infty}\left[H\left(\De\right)-H\left(-\De\right)\right]\frac{d\De}{\Delta\epsilon^{2}}\right)\,,\label{eq:DC1-iso-inf}\\
\Gamma_{2}^{\Lambda}\left(E\right) & = & \mathcal{K}\left(E\right)\left(\int_{\Demin}^{\infty}\left[H\left(\De\right)+H\left(-\De\right)\right]\frac{d\De}{\Delta\epsilon}\right)\,.\label{eq:DC2-iso-inf}
\end{eqnarray}
After integrating by parts, taking the limit $\Demin\to0$ and using
the relations $\Gamma_{2}\left(E\right)=\mathcal{K}_{E}H\left(0\right)$
and $\Gamma_{1}\left(E\right)=\mathcal{K}_{E}H^{\prime}\left(0\right)$,
the DCs can be written as 

\begin{eqnarray}
\Gamma_{1}^{\Lambda}\left(E\right) & \approx & \Gamma_{1}\left(E\right)\log\left(e^{\left\{ H\left(-1\right)-H\left(1\right)-\int_{0}^{1}\left[H^{\prime\prime}\left(\De\right)-H^{\prime\prime}\left(-\De\right)\right]\log\De d\De+\int_{1}^{\infty}\left[H\left(-\De\right)+H\left(\De\right)\right]d\De/\Delta\epsilon^{2}\right\} /2H^{\prime}\left(0\right)}/\Demin\right)\,,\label{eq:DC1-iso-inf-eff}\\
\Gamma_{2}^{\Lambda}\left(E\right) & \approx & \Gamma_{2}\left(E\right)\log\left(e^{\left\{ -\int_{0}^{1}\left[H^{\prime}\left(\De\right)-H^{\prime}\left(-\De\right)\right]\log\De d\De+\int_{1}^{\infty}\left[H\left(-\De\right)+H\left(\De\right)\right]d\De/\Delta\epsilon\right\} /2H\left(0\right)}/\Demin\right)\,.\label{eq:DC2-iso-inf-eff}
\end{eqnarray}
Note that for a power law distribution, $g\left(s\right)=s^{p}$,
Eq. \eqref{eq:H-de} reads 
\begin{eqnarray}
H\left(\De\right)=\begin{cases}
\frac{1}{2\sqrt{\pi}}\left(-\De\right)^{p-1/2}\left(5-2p\right)\Gamma\left(\frac{1}{2}-p\right)\Gamma\left(1+p\right) & \De<-1\\
\frac{1}{4}\left[\frac{(2-2p-3\De)(1+\De)^{1+p}}{1+p}+(5-2p)(-\De)^{-\frac{1}{2}+p}\beta\left(-\De,\frac{1}{2}-p,1+p\right)\right] & -1<\De<0\\
\frac{3}{4(1+p)}\frac{4\left(3-2p\right)+\left[13-2p(7-4p)\right]\De}{3-4(2-p)p}(1+\De)^{p-3/2} & \De>0
\end{cases}
\end{eqnarray}
where $\beta\left(x;a,b\right)$ is the incomplete beta function.

\section{Evaluation of the Strong Propagator}

\label{s:propagator}

We derive here a perturbative expression for the SP in the limit $\DEmax\to\infty$.
In this limit, which is the relevant one for evolution on short timescales,
the problem is closely related to the problem of Pareto sums with
index $\alpha=2$. 

Consider a star with binding energy $E$ and angular momentum $J$.
The orbit-averaged rate with which the test star changes its energy
is given by the transition probability (Eq. \ref{eq:K-JE}) 
\begin{equation}
K_{E,J}(\DE)=\frac{\Gamma_{2}(E,J)+\Gamma_{1}(E,J)\DE}{2\left|\DE\right|^{3}}\,.\label{eq:K-JE-App}
\end{equation}
When the timescale of interest is much shorter than the energy relaxation
time, $\Gamma_{1}$ and $\Gamma_{2}$ can be approximated as constant
in $E$ and $J$ and the maximal energy change in a single scattering
event is much smaller than the large energy exchange cutoff $\Delta E\ll\Delta E_{\max}$.
Therefore $\Delta E_{\max}$ can be taken to the limit $\Delta E_{\max}\to\infty$.
In that limit the Fourier transform of $K_{E,J}(\DE)$ is given by
\begin{eqnarray}
\tilde{K}_{E,J}(k) & = & \int_{\DEmin}^{\infty}e^{-ik\Delta E}K_{E.J}\left(\Delta E\right)\nonumber \\
 & = & \frac{\Gamma_{2}(E,J)}{2\DEmin^{-2}}\left\{ \cos(k\DEmin)+(k\DEmin)^{2}Ci\left(k\DEmin\right)-k\DEmin\sin(k\DEmin)\right.\nonumber \\
 &  & \left.+2\eta\De_{\min}i\left[\sin(k\DEmin)-(k\DEmin)Ci(k\DEmin)\right]\right\} \nonumber \\
 & = & q_{2b}\left\{ 1+\eta i\left(2-2\gamma_{E}-\log(k^{2}\DEmin^{2})\right)(k\DEmin)\right.\nonumber \\
 &  & \left.-\frac{1}{2}\left(3-2\gamma_{E}-\log(k^{2}\DEmin^{2})\right)(k\DEmin)^{2}+O\left(k^{3}\DEmin^{3}\right)\right\} \,,\label{eq:K-FT-App}
\end{eqnarray}
where $\eta\De_{\min}=E\Gamma_{1}(E,J)/\Gamma_{2}(E,J)\De_{\min}\ll1$,
$\De_{\min}=\DEmin/E$, $q_{2b}=\int K_{E,J}\left(\DE\right)d\DE$
and $\gamma_{E}$ is the Euler's gamma constant. The energy propagator
can therefore be directly obtained from Eq. \eqref{eq:W-FT} 
\begin{eqnarray}
W(\DE,t) & = & \frac{1}{2\pi}\int_{-\infty}^{\infty}e^{-ik\Delta E}dk\nonumber \\
 &  & \exp\left\{ \eta\De_{\min}i\left(2-2\gamma_{E}-\log(k^{2}\DEmin^{2})\right)k\DEmin q_{2b}t\right.\nonumber \\
 &  & \left.-\frac{1}{2}\left(3-2\gamma_{E}-\log(k^{2}\DEmin^{2})\right)(k\DEmin)^{2}q_{2b}t\right\} \,.\label{eq:W-FT-inf}
\end{eqnarray}
Define $\tau=q_{2b}t$, $\sigma(\tau)=\sqrt{2\tau\log L(\tau)\DEmin^{2}}$
and $\mu(\tau)=2\eta\De_{\min}\tau\log L(\tau)\DEmin$, where $L(\tau)$
is defined by
\begin{eqnarray}
2\log L\left(\tau\right)-\log(2\log L\left(\tau\right)) & = & 3-2\gamma_{E}+\log(\tau)\,,\label{eq:L-t-def}
\end{eqnarray}
and
\begin{eqnarray}
L\left(\tau\right) & = & \exp\left\{ -\frac{1}{2}W_{-1}\left(-\frac{e^{-3+2\gamma_{E}}}{\tau}\right)\right\} \approx\sqrt{e^{3-2\gamma_{E}}\tau}\exp\left\{ 1+\frac{\log\log\left(e^{3-2\gamma_{E}}\tau\right)}{\log\left(e^{3-2\gamma_{E}}\tau\right)}\right\} \,,\label{eq:L-t-app}
\end{eqnarray}
where $W_{-1}(\tau)$ is the Lambert W function, and $L$ is an increasing
function of time. By a change of variables to $x=(\DE-\mu(\tau))/\sigma(\tau)$
and $\omega=k\sigma(\tau)$, Eq. (\ref{eq:W-FT-inf}) reads
\begin{eqnarray}
W\left(x,\tau\right) & \approx & \frac{1}{2\pi}\int_{-\infty}^{\infty}d\omega e^{-i\omega x}e^{-\frac{1}{2}\omega^{2}}\exp\left\{ \frac{\omega^{2}\log\omega^{2}}{4\log L\left(\tau\right)}-i\eta\De_{\min}\left(\sqrt{\frac{\tau}{2\log L\left(\tau\right)}}\left(1+\log\omega^{2}\right)\right)\omega\right\} \nonumber \\
 & \approx & \frac{1}{2\pi}\int_{-\infty}^{\infty}d\omega e^{-i\omega x}e^{-\frac{1}{2}\omega^{2}}\left\{ 1+\frac{\omega^{2}\log\omega^{2}}{4\log L\left(\tau\right)}-i\eta\De_{\min}\left(\sqrt{\frac{\tau}{2\log L\left(\tau\right)}}\left(1+\log\omega^{2}\right)\right)\omega\right\}\nonumber \\
 & \approx & \frac{1}{\sqrt{2\pi}}e^{-\frac{x^{2}}{2}}+\frac{1}{4\log L\left(\tau\right)}\left(W_{c}^{1}(x)+2\eta\De_{\min}\frac{\mu(\tau)}{\sigma(\tau)}W_{c}^{2}(x)\right)\,,\label{eq:W-x-t}
\end{eqnarray}
where 
\begin{eqnarray}
W_{c}^{1}(x) & \equiv & \mathcal{F}^{-1}\left\{ e^{-\frac{\omega^{2}}{2}}\omega^{2}\log(\omega^{2})\right\} =\frac{1}{\sqrt{2\pi}}e^{-\frac{x^{2}}{2}}\left\{ 2+\left(x^{2}-1\right)\left[F_{2}(x^{2})+\gamma_{E}+\log2\right]\right\} -x\mathrm{erf}\left(\frac{x}{\sqrt{2}}\right)\label{eq:Wc1}
\end{eqnarray}
and 
\begin{eqnarray}
W_{c}^{2}(x) & \equiv & \mathcal{F}^{-1}\left\{ -ie^{-\frac{\omega^{2}}{2}}\omega(1+\log(\omega^{2}))\right\} =\frac{1}{\sqrt{2\pi}}e^{-\frac{x^{2}}{2}}x\left(\gamma_{E}-1+\log2+F_{2}(x^{2})\right)-\mathrm{erf}\left(\frac{x}{\sqrt{2}}\right)\,,\label{eq:Wc2}
\end{eqnarray}
where $F_{2}(x)\equiv x{}_{2}F_{2}(1,1,3/2,2,x/2)$ and $_{2}F_{2}$
is a generalized Hypergeometric function. 

Integrating Eqs. \eqref{eq:Wc1}, \eqref{eq:Wc2} to obtain 
\begin{eqnarray}
FW_{c}^{1}\left(x\right)=\int_{-\infty}^{x}W_{c}^{1}(x^{\prime})dx^{\prime} & = & \mathcal{F}^{-1}\left\{ i\omega e^{-\frac{\omega^{2}}{2}}\log(\omega^{2})\right\} =\mathrm{erf}\left(\frac{x}{\sqrt{2}}\right)-\frac{1}{\sqrt{2\pi}}e^{-\frac{x^{2}}{2}}x\left(\gamma_{E}+\log2+F_{2}(x^{2})\right)\,,\label{eq:FWc1}
\end{eqnarray}
and 
\begin{eqnarray}
FW_{c}^{2}\left(x\right)=\int_{-\infty}^{x}W_{c}^{2}(x^{\prime})dx^{\prime} & = & \mathcal{F}^{-1}\left\{ e^{-\frac{\omega^{2}}{2}}(1+\log(\omega^{2}))\right\} =-\frac{1}{\sqrt{2\pi}}e^{-\frac{x^{2}}{2}}\left[\gamma_{E}-1+\log2+F_{2}(x^{2})\right]\label{eq:FWc2}
\end{eqnarray}
and therefore
\begin{eqnarray*}
W\left(X<x\right) & \approx & \frac{1}{2}\left[1+\mathrm{erf}\left(\frac{x}{\sqrt{2}}\right)\right]+\frac{1}{4\log L\left(\tau\right)}\left(FW_{c}^{1}(x)+2\eta\De_{\min}\frac{\mu(\tau)}{\sigma(\tau)}FW_{c}^{2}(x)\right)
\end{eqnarray*}
By Defining 
\begin{equation}
\Gamma_{1,2}^{L}(E,J,t)\equiv\Gamma_{1,2}(E,J)\log L\left(q_{2b}\left(E,J\right)t\right)\,,\label{eq:Gamma-tild-app}
\end{equation}
we can write $\mu\left(t\right)=t\Gamma_{1}^{L}(E,J,t)$ and $\sigma^{2}\left(t\right)=t\Gamma_{2}^{L}(E,J,t)$
and Eq. (\ref{eq:W-x-t}) reads 
\begin{eqnarray}
W_{E,J}\left(\DE,t\right) & \approx & \frac{1}{\sqrt{2\pi t\Gamma_{2}^{L}(E,J,t)}}e^{-\frac{\left(\DE-t\Gamma_{1}^{L}(E,J,t)\right)^{2}}{2t\Gamma_{2}^{L}(E,J,t)}}+\frac{1}{4\log L\left(q_{2b}t\right)}\frac{1}{\sqrt{t\Gamma_{2}^{L}(E,J,t)}}W_{c}^{1}\left(\frac{\DE-t\Gamma_{1}^{L}(E,J,t)}{\sqrt{t\Gamma_{2}^{L}(E,J,t)}}\right)\nonumber \\
 &  & +\frac{1}{\sqrt{4\log L\left(q_{2b}t\right)}}\frac{1}{\sqrt{t\Gamma_{2}^{L}(E,J,t)}}\eta\sqrt{\frac{\Gamma_{2}(E,J)t}{E^{2}}}W_{c}^{2}\left(\frac{\DE-t\Gamma_{1}^{L}(E,J,t)}{\sqrt{t\Gamma_{2}^{L}(E,J,t)}}\right)\,.\label{eq:W-DE}
\end{eqnarray}
This is a modification to the impulse response of the FP propagator
(Eq. \eqref{eq:W-FP}) that is a Gaussian with a variance of $t\Gamma_{2}\log\Lambda$.
If we assume that $\DEmax\sim E$ and $\DEmin\sim\Lambda^{-1}E$,
then the time $t_{\Lambda}$ for which the Gaussian in Eq. \eqref{eq:W-DE}
has the same width of the FP Gaussian is 
\begin{equation}
t_{\Lambda}=e^{2\gamma_{E}-3}\frac{E^{2}}{\Gamma_{2}\log\Lambda}\approx t_{E}/6\,.\label{eq:t-Lambda-app}
\end{equation}

To the first order, the asymptotic expansions of $W_{c}^{1}$ and
$W_{c}^{2}$ are $x/|x|^{3}$, $2/|x|^{3}$ respectively, and thus
all the moments diverge. The mean and variance are therefore not useful
estimators. However, the fractional moments (i.e. $\mu_{\delta}\equiv\langle|\DE|^{\delta}\rangle$
where $\delta<1$) do converge, which suggest that these moments can
be used as estimators. Using Eq. (\ref{eq:W-DE}) the fractional moments
is defined as 
\begin{eqnarray}
\mu_{\delta}(\tau) & \equiv & \int d\DE\left|\DE\right|^{\delta}W_{E,J}\left(\DE,t\right)\nonumber \\
 & = & (\sqrt{2}\sigma)^{\delta}\frac{1}{\sqrt{\pi}}\Gamma\left(\frac{1+\delta}{2}\right){}_{1}F_{1}\left(-\frac{\delta}{2},\frac{1}{2},-\frac{1}{2}\frac{\mu^{2}}{\sigma^{2}}\right)\,\nonumber \\
 &  & +\frac{\sigma^{\delta}}{4\log L\left(\tau\right)}\int\left|x\right|^{\delta}\left[W_{c}^{1}\left(x-\frac{\mu}{\sigma}\right)+\eta\frac{\mu}{\sigma}\De_{\min}W_{c}^{2}\left(x-\frac{\mu}{\sigma}\right)\right]dx\,.
\end{eqnarray}
In the limit $\delta\ll1$ and$\mu^{2}/\sigma^{2}\ll1$
\begin{eqnarray}
\mu_{\delta}^{2/\delta} & \approx & 2\sigma^{2}\left(\frac{1}{\sqrt{\pi}}\Gamma\left[\frac{1+\delta}{2}\right]\right)^{2/\delta}\,,\label{eq:mu-delta-exp}
\end{eqnarray}
and therefore $\sigma^{2}$ can be estimated using 
\begin{equation}
\sigma^{2}=\lim_{\delta\to0}\frac{1}{2}\left(\frac{\left\langle \DE\right\rangle }{\Gamma\left(\frac{1+\delta}{2}\right)/\sqrt{\pi}}\right)^{2/\delta}\,.\label{eq:sigma-delta}
\end{equation}

\section{Time-dependent solutions of the FP equation for a power-law cusp}

\label{s:FP-solution}

\global\long\def\Np{N_{p}}

\global\long\def\Fp{F_{p}}

\global\long\def\Er{E_{h}}

\global\long\def\tE{t_{E}^{h}}

\global\long\def\xin{x_{0}}

We solve Here analytically the FP equation describing the evolution
of a small perturbation on a fixed background distribution of a power-law
cusp around a MBH. Consider a small perturbation $\Np(E)$ on a fixed
background distribution $N_{b}(E)\propto E^{p-5/2}$. Assuming an
infinite cusp, the DCs are given by $\Gamma_{1}^{\Lambda}(E)\propto E^{p+1}$
and $\Gamma_{2}^{\Lambda}(E)\propto E^{p+2}$. Define $x=E/\Er$,
where $\Er$ is some reference energy, for example $\Er=\sigma_{0}^{2}$
where $\sigma_{0}$ is the velocity dispersion far from the MBH.

The FP equation for the small perturbation $N_{p}$ is 
\begin{equation}
\frac{\partial\Np\left(x,\tau\right)}{\partial\tau}=\frac{1}{2}\frac{\partial}{\partial x^{2}}x^{p+2}\Np\left(x,\tau\right)-\eta\frac{\partial}{\partial x}x^{p+1}\Np\left(x,\tau\right)=\frac{\partial\Fp}{\partial x}\,,\label{eq:FP-x}
\end{equation}
where $\eta=\Er\Gamma_{1}^{\Lambda}(\Er)/\Gamma_{2}^{\Lambda}(\Er)=-(1-4p)(1-2p)/(4(3-2p))$
and $\tau=t/\tE$ where $\tE=\Er^{2}/(\Gamma_{2}(\Er)\log\Lambda)$.
The steady state solution (assuming $p+2-2\eta>0$) is $\Np^{\infty}(E)\propto E^{2\eta-p-2}$.
Assuming particles cannot leave the system and are restricted to energies
$E>\Er$, the steady state is given by $\Np^{\infty}\left(x\right)=\left(p+1-2\eta\right)x^{2\eta-p-2}$.

The time dependent solution can be obtained by the eigenfunction method
\citep[e.g.][]{Gardiner2004}. Let $\Np(x,\tau)=\Np^{\infty}(x)q(x,\tau)$.
We are looking solutions of the form $\Np(x,\tau)=P_{\lambda}(x)e^{-\lambda\tau}$
and $q(x,\tau)=Q_{\lambda}(x)e^{-\lambda\tau}$, which satisfy 
\begin{eqnarray}
\frac{1}{2}\frac{\partial}{\partial x^{2}}\{x^{p+2}P_{\lambda}(x)\}-\eta\frac{\partial}{\partial x}\{x^{p+1}P_{\lambda}(x)\} & = & -\lambda P_{\lambda}(x)\,,\label{eq:ign-func-equ}\\
\frac{1}{2}x^{p+2}\frac{\partial^{2}}{\partial x^{2}}Q_{\lambda}(x)+\eta x^{p+1}\frac{\partial}{\partial x}Q_{\lambda}(x) & = & -\lambda Q_{\lambda}(x)\,,\label{eq:ign-func-equ-Q}
\end{eqnarray}
and 
\begin{equation}
\Fp(x)=\frac{p+1-2\eta}{2}x^{2\eta}e^{-\lambda\tau}\frac{\partial}{\partial x}Q(x)\,.\label{eq:F(x)}
\end{equation}
For $p>0$ the solutions of Eqs. (\eqref{eq:ign-func-equ},~\eqref{eq:ign-func-equ-Q})
is of the form 
\begin{eqnarray}
P_{\lambda}\left(x\right) & = & C_{\lambda}x^{\eta-p-3/2}J_{\alpha}\left(\frac{2\sqrt{2\lambda}}{px^{p/2}}\right)\,,\\
Q_{\lambda}\left(x\right) & = & \frac{C_{\lambda}}{p+1-2\eta}x^{1/2-\eta}J_{\alpha}\left(\frac{2\sqrt{2\lambda}}{px^{p/2}}\right)\,,\label{eq:P(x)}
\end{eqnarray}
and 
\begin{equation}
\Fp(x)=\frac{C_{\lambda}2\sqrt{2\lambda}}{4}x^{-(1+p+2h)/2}J_{\alpha+1}\left(\frac{2\sqrt{2\lambda}}{px^{p/2}}\right)\,,\label{eq:F(x)-explicit}
\end{equation}
where $\alpha=(1-2\eta)/p$, and $J_{v}(x)$ is the Bessel function
of the first kind, and where we assumed $\Fp(x)=0$ as $x\to\infty$.
The assumption that stars cannot leave the system and are restricted
to energies $E>\Er$ is expressed by a reflecting boundary at $\Er$,
i.e. $\Fp(x=1)=0$. The eigenvalues are therefore 
\begin{equation}
\lambda_{n}=\frac{p^{2}}{8}j_{\alpha+1,n}^{2}\,,\label{eq:lambda-n}
\end{equation}
where $j_{v,n}$ is the n-th zero of $J_{v}(x)$.

By choosing $C_{\lambda}=\frac{\sqrt{p(p+2-2\eta)}}{J_{\alpha}(j_{\alpha+1,n})}$,
the eigenfunctions $Q_{n}\equiv Q_{\lambda_{n}}$ and $P_{n}\equiv P_{\lambda_{n}}$
satisfy the bi-orthogonal relations

\begin{equation}
\int_{1}^{\infty}P_{n}(x)Q_{m}(x)=\frac{2}{J_{\alpha}(j_{\alpha+1,n})J_{\alpha}(j_{\alpha+1,m})}\int_{0}^{1}sJ_{\alpha}(j_{\alpha+1,n}s)J_{\alpha}(j_{\alpha+1,m}s)ds=\delta_{n,m}\,.\label{eq:PQ-orthogonality}
\end{equation}
Any solution can therefore be written in the form

\begin{equation}
\Np(x,t)=\sum_{n}A_{n}P_{n}(x)e^{-\lambda_{n}t}\,,\label{eq:N(x,t)-sum}
\end{equation}
where $A_{n}=\int_{1}^{\infty}Q_{n}\Np(x,0)$.

The time dependent distributions resulting for initial conditions
of $\Np(x,0)=\delta(x-\xin)$, for example, are given by 
\begin{equation}
\Np\left(x,t|\xin,0\right)=\Np^{\infty}(x)\left[1+\sum_{n}Q_{n}(x)Q_{n}(\xin)e^{-\frac{p^{2}}{8}j_{\alpha+1,n}^{2}\tau}\right]\,,\label{eq:N(x,t)-impulse-respond}
\end{equation}
where 
\begin{equation}
Q_{n}\left(x\right)=\frac{x^{1/2-\eta}}{J_{\alpha}(j_{\alpha+1,n})}\sqrt{\frac{p}{p+1-2\eta}}J_{\alpha}\left(j_{\alpha+1,n}x^{-p/2}\right)\,.\label{eq:Q(x)-normalized}
\end{equation}

A different situation is obtained when the test stars can leave the
system. This can be approximated by imposing a boundary condition
$\Np(x=1,t)=0$. In that case $\lambda=p^{2}j_{\alpha,n}^{2}/8$ and
\begin{equation}
Q\left(x\right)=\sqrt{\frac{p}{p+1-2\eta}}\frac{x^{1/2-\eta}}{J_{\alpha+1}\left(j_{\alpha,n}\right)}J_{\alpha}\left(j_{\alpha,n}x^{-p/2}\right)\,.\label{eq:Q(x)-absorb}
\end{equation}
The response for $\delta(x-\xin)$ initial conditions is then given
by 
\begin{equation}
\Np\left(x,t|\xin,0\right)=\Np^{\infty}\left(x\right)\sum_{n}Q_{n}(x)Q_{n}(\xin)e^{-\frac{p^{2}}{8}j_{\alpha,n}^{2}\tau}\,,\label{eq:N(x,t)-impulse-respond-absorb}
\end{equation}
where the number of test stars with relative energy larger than $x$
at time $\tau$ is given by 
\begin{equation}
\Np(>x,\tau)=2\left(x/\xin\right)^{1/2-\eta}x^{-p/2}\sum_{n}\frac{J_{\alpha+1}\left(j_{\alpha,n}x^{-p/2}\right)J_{\alpha}\left(j_{\alpha,n}\xin^{-p/2}\right)}{j_{\alpha,n}J_{\alpha+1}^{2}\left(j_{\alpha,n}\right)}e^{-\frac{p^{2}}{8}j_{\alpha,n}^{2}\tau}\,.\label{eq:N-p(X,t)}
\end{equation}

\section{Energy DCs for a finite isotropic power-law cusp}

\label{s:cusp-solution}

Cusps around MBHs, whether real or in $N$-body simulations, contain
a finite number of stars, and are therefore finite in extent. As we
show here, these edges can substantially increase the relaxation timescale. 

Consider a system with $N$ stars of mass $m$ each, with a semi-major
axis distribution $\rho(a)da\propto a^{2-\gamma}da$ and eccentricity
distribution $\rho(e)de=2ede$. The number of stars with energy larger
than $E$ is $N(>E)=N_{tot}(E_{\min}/E)^{3-\gamma}$ where $N_{tot}$
is the total number of stars up to energy $E_{\min}$ and $\gamma=3/2+p$.
Since the number of stars in the system is finite, the DF is truncated
at some maximal energy $E_{\max}\approx N_{tot}^{1/(3-\gamma)}E_{\min}$,
where $N(<E_{\max})=1$.

Eqs. \eqref{eq:Gamma1-iso}, \eqref{eq:Gamma2-iso} imply (for $0.5<\gamma<2$)
\begin{eqnarray}
\Gamma_{1}^{\Lambda}(E) & = & \log\Lambda\frac{E}{P(E)}\frac{c_{1}}{Q^{2}}N(>E)S_{E1}(E,E_{\min},E_{\max})\label{eq:Gamma1-iso-cusp}
\end{eqnarray}
and
\begin{eqnarray}
\Gamma_{2}^{\Lambda}(E) & = & \log\Lambda\frac{E^{2}}{P(E)}\frac{c_{2}^{2}}{Q^{2}}N(>E)S_{E22}(E,E_{\min},E_{\max})\,,\label{eq:Gamma2-iso-cusp}
\end{eqnarray}
where 
\begin{equation}
c_{1}=\frac{4(1-4p)}{1+p}\,,\label{eq:alpha1}
\end{equation}
 
\begin{equation}
c_{2}^{2}=16\frac{3-2p}{(1+p)(1-2p)}\,,\label{eq:alpha2}
\end{equation}
and $S_{E1}$, $S_{E2}$ are the correction functions due to the edges,
whose parameters are $p$ and $N_{tot}$ (via $E_{\max}/E_{\min}=N_{tot}^{1/(3/2-p)}$),
\begin{equation}
S_{E1}(E)=\begin{cases}
-\frac{2+2p}{1-4p}\left(1-\left(\frac{E_{\min}}{E_{\max}}\right)^{3/2-p}\right)\left(\frac{E}{E_{\min}}\right)^{3/2-p} & E<E_{\min}\\
1-\frac{3-2p}{1-4p}\left(\frac{E_{\min}}{E}\right)^{p+1}+\frac{2+2p}{1-4p}\left(\frac{E}{E_{\max}}\right)^{3/2-p} & E_{\min}>E>E_{\max}\\
\frac{3-2p}{1-4p}\left(1-\left(\frac{E_{\min}}{E_{\max}}\right)^{1+p}\right)\left(\frac{E_{\max}}{E}\right)^{1+p} & E>E_{\max}
\end{cases}\label{eq:S-E1}
\end{equation}
and 
\begin{equation}
S_{E22}(E)=\begin{cases}
\frac{2+2p}{3}\left(\frac{E}{E_{\min}}\right)^{1/2-p}\left(1-\left(\frac{E_{\min}}{E_{\max}}\right)^{1/2-p}\right) & E>E_{\min}\\
1-\frac{1-2p}{3}\left(\frac{E_{\min}}{E}\right)^{p+1}-\frac{2+2p}{3}\left(\frac{E}{E_{\max}}\right)^{1/2-p} & E_{\min}>E>E_{\max}\\
\frac{1-2p}{3}\left(\frac{E_{\max}}{E}\right)^{p+1}\left(1-\left(\frac{E_{\min}}{E_{\max}}\right)^{p+1}\right) & E>E_{\max}
\end{cases}\,.\label{eq:S-E22}
\end{equation}

\begin{figure}
\centering \includegraphics[width=0.5\columnwidth]{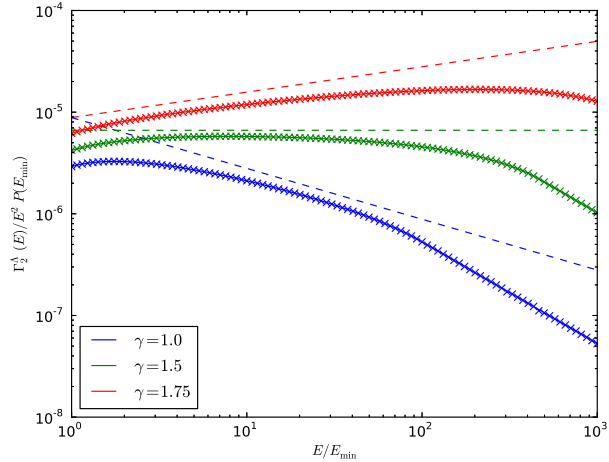}
\caption{\label{fig:Relaxation-analytic} Energy relaxation as function of
energy, for a cusp ($N(>E)\propto E^{3-\gamma}$) with $N_{tot}=10^{4}$
stars and $Q=10^{6}$, for different power-law slopes. The solid lines
are the relaxation rates calculated by Eq. (\ref{eq:Gamma2-iso-cusp}),
the crosses are the rates calculated using the \citet{Cohn1978} integrals,
the dashed lines are the rates for infinite cusps.}
\end{figure}

Figure~\ref{fig:Relaxation-analytic} shows that the truncation of
the DF in a finite cusp leads to slower relaxation.

\bibliographystyle{apj}

\end{document}